\documentclass[12pt,preprint]{aastex}
\usepackage{emulateapj5}
\usepackage{apjfonts}
\usepackage{epsfig}

\newcommand{\chandra}{\emph{Chandra}}

\newcommand{\etal}{et al.}

\def\gtrsim{\mathrel{\hbox{\rlap{\hbox{\lower4pt\hbox{$\sim$}}}\hbox{\raise2pt\hbox{$>$}}}}}

\newcommand{\fwoiii}{\ensuremath{\mathrm{FWHM}_\mathrm{[O \tiny III]}}}

\newcommand{\hbeta}{H\ensuremath{\beta}}

\newcommand{\hst}{\emph{HST}}

\newcommand{\kms}{km~s\ensuremath{^{-1}}}

\newcommand{\lledd}{\ensuremath{L_{\mathrm{bol}}/L{\mathrm{_{Edd}}}}}

\newcommand{\loiii}{\ensuremath{L_{\mathrm{[O {\tiny III}]}}}}

\newcommand{\msun}{\ensuremath{M_{\odot}}}

\newcommand{\oii}{[\ion{O}{2}]}
\newcommand{\oiii}{[\ion{O}{3}]}

\newcommand{\sii}{[\ion{S}{2}]}

\newcommand{\sigmastar}{\ensuremath{\sigma_{\ast}}}
\newcommand{\sigmagas}{\ensuremath{\sigma_{\rm g}}}

\newcommand{\spitzer}{\emph{Spitzer}}

\newcommand{\xmm}{{\it XMM-Newton}}

\def\lax{{$\mathrel{\hbox{\rlap{\hbox{\lower4pt\hbox{$\sim$}}}\hbox{$<$}}}$}}
\def\gax{{$\mathrel{\hbox{\rlap{\hbox{\lower4pt\hbox{$\sim$}}}\hbox{$>$}}}$}}

\slugcomment{Feb 14, 2011; accepted for publication in {\it The Astrophysical Journal}.}
\shorttitle{{\it T2 QSOs}}
\shortauthors{GREENE, ET AL.}

\begin{document}

\title{Feedback in Luminous Obscured Quasars}

\author{Jenny E. Greene}
\affil{Department of Astronomy, UT Austin, 1 University Station C1400, 
Austin, TX 71712}

\author{Nadia L. Zakamska}
\affil{Kavli Institute for Particle Astrophysics and Cosmology,
  Stanford University, 2575 Sand Hill Road, MS-29, Menlo Park, CA 94025; Kavli Fellow}

\author{Luis C. Ho}
\affil{The Observatories of the Carnegie Institution for Science,
813 Santa Barbara Street, Pasadena, CA 91101}

\author{Aaron J. Barth}
\affil{Department of Physics and Astronomy, 4129 Frederick Reines Hall, 
University of California, Irvine, CA 92697-4575}

\begin{abstract}

We use spatially resolved long-slit spectroscopy from Magellan to
investigate the extent, kinematics, and ionization structure in the
narrow-line regions of 15 luminous, obscured quasars with
$z<0.5$. Increasing the dynamic range in luminosity by an order of
magnitude, as well as improving the depth of existing observations
by a similar factor, we revisit relations between narrow-line region
size and the luminosity and linewidth of the narrow emission
lines. We find a slope of $0.22\pm0.04$ for the power-law
relationship between size and luminosity, suggesting that the 
nebulae are limited by availability of gas to ionize at these luminosities.  In fact, we find that
the active galactic nucleus is effectively ionizing the
interstellar medium over the full extent of the host galaxy. Broad
($\sim 300-1000$ km~s$^{-1}$) linewidths across the galaxies reveal
that the gas is kinematically disturbed.  Furthermore, the rotation
curves and velocity dispersions of the ionized gas remain constant
out to large distances, in striking contrast to normal and starburst
galaxies. We argue that the gas in the entire host galaxy is significantly
disturbed by the central active galactic nucleus. While only $\sim
10^7-10^8$~\msun\ worth of gas are directly observed to be leaving
the host galaxies at or above their escape velocities, these
estimates are likely lower limits because of the biases in both mass
and outflow velocity measurements and may in fact be in accord with
expectations of recent feedback models. Additionally, we report the
discovery of two dual obscured quasars, one of which is blowing a
large-scale ($\sim 10$ kpc) bubble of ionized gas into the
intergalactic medium.

\end{abstract}

\section{Introduction}

It has become fashionable to invoke feedback from accreting black
holes (BHs) as an influential element of galaxy evolution
\citep[e.g.,][]{taborbinney1993,silkrees1998,hopkinsetal2006,
  sironisocrates2010}. Regulatory mechanisms are sorely needed to keep
massive galaxies from forming too many stars and becoming overly
massive or blue at late times
\citep[e.g.,][]{thoulweinberg1995,crotonetal2006}.  Feedback from an
accreting BH provides a tidy solution.  For one thing, the
gravitational binding energy of a supermassive BH is completely
adequate to unbind leftover gas in the surrounding
galaxy. Furthermore, using simple prescriptions for black hole
feedback leads to a natural explanation for the observed scaling
relations between the BH mass and properties of the surrounding
galaxy, including stellar velocity dispersion and bulge luminosity and
mass \citep[e.g.,][]{haeringrix2004,gultekinetal2009}.  The problem
remains to find concrete evidence of BH self-regulation, and to
determine whether or not accretion energy has a direct impact on the
surrounding galaxy.

There are some special circumstances in which accretion energy clearly
has had an impact on its environment.  For instance, jet activity in
massive elliptical galaxies and brightest cluster galaxies deposits
energy into the hot gas envelope \citep[see review in
][]{mcnamaranulsen2007}, although the efficiency of coupling the
accretion energy to the gas remains uncertain
\citep[e.g.,][]{birzanetal2004,bestetal2005,heinzetal2006}, as does
the relative importance of heating by the active nucleus as opposed to
other possible sources \citep[e.g.,][]{zakamskanarayan2003,sijackietal2008,
conroyostriker2008,parrishetal2009}.
Likewise, there is clear evidence that powerful radio jets entrain
warm gas and carry significant amounts of material out of their host
galaxies \citep[e.g.,][]{vanbreugeletal1986,tadhunter1991,whittle1992iii,
  villarmartinetal1999,nesvadbaetal2006,nesvadbaetal2008}.  However,
as only a minority ($\sim 10\%$) of active BHs are radio-loud,
invoking this mechanism as the primary mode of BH feedback would require
all galaxies to have undergone a radio-loud phase -- a conjecture
which lacks direct evidence and contradicts a theoretical paradigm in
which radio-loudness is determined by the spin of the black hole
\citep[e.g.,][]{tchekhovskoyetal2010}. Thus, it is not clear whether
BH activity in radio galaxies accounts for more than a small fraction
of the BH growth \citep[e.g.,][]{soltan1982,yutremaine2002} and
therefore whether this mode of feedback is in fact the dominant
one. 

Nuclear activity is known to drive outflows on small scales.  Broad
absorption-line troughs are seen in $\sim 10-20\%$ of luminous quasars
\citep[e.g,][]{reichardetal2003}, and there is good reason to believe
that the outflows are ubiquitous but have a covering fraction of $\sim
20\%$, at least for high-\lledd\ systems
\citep[e.g.,][]{weymannetal1981,gallagheretal2007,shenetal2008bal}.
The velocities in broad absorption lines are high ($\sim
10,000$~\kms), and they most likely emerge from a wind blown off of
the accretion disk \citep[e.g.,][]{progakallman2004}. In a few rare
objects the outflow appears to extend out to large distances from the
nucleus \citep[][]{moeetal2009}, but it is unclear whether most of
these outflows have any impact beyond hundreds of Schwarzschild
radii. Narrow associated absorption-line systems are signposts of
outflows extending to larger distances, but determining 

\vbox{ 
\hskip 0.in
\psfig{file=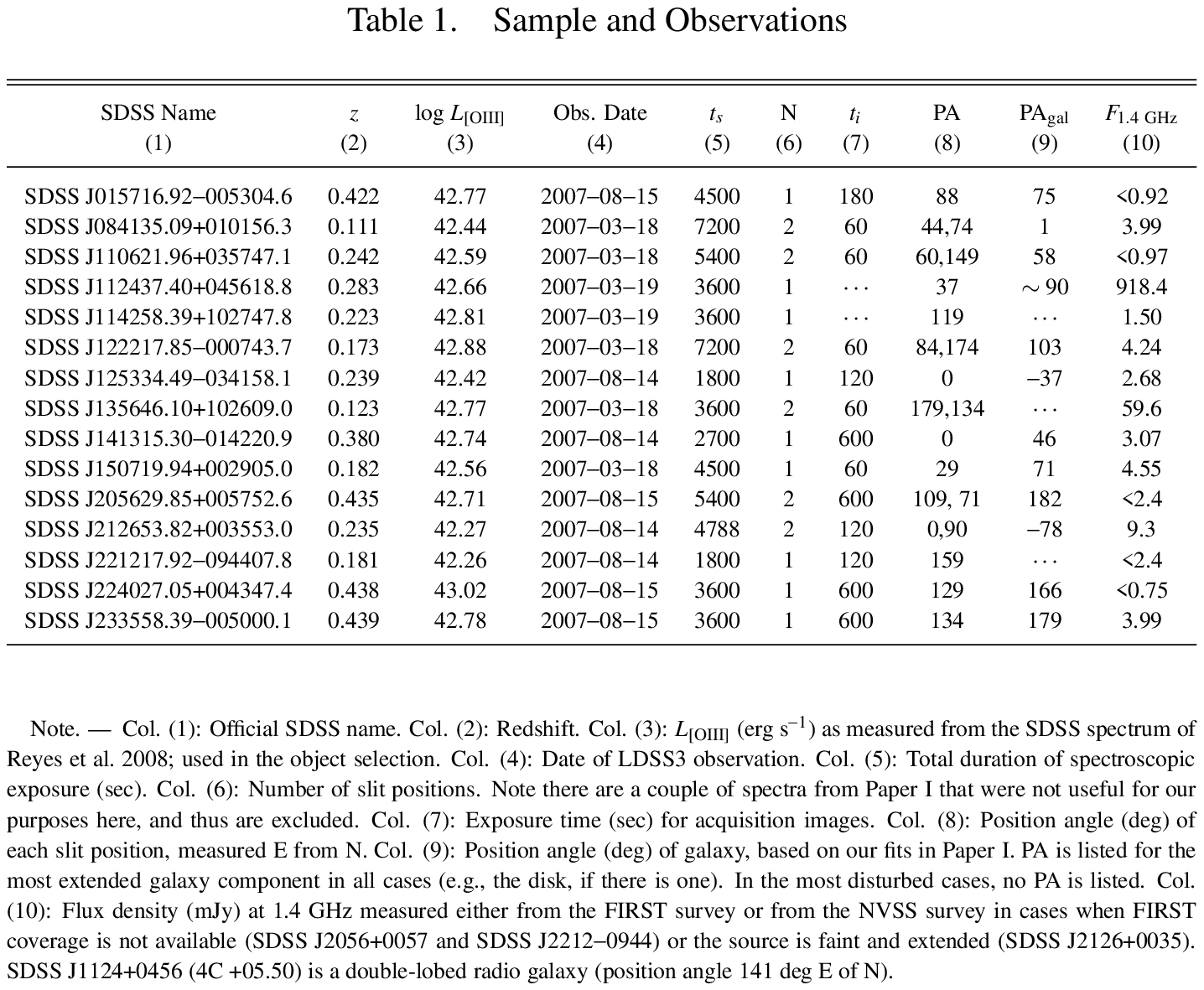,width=0.45\textwidth,keepaspectratio=true,angle=0}
}
\vskip 4mm
\noindent
their physical
radii (and thus the mass outflow rate) is notoriously challenging.  In
cases where it is possible, the estimated outflow rates are thought to
be significant fractions of the accretion onto the BH \citep[see
review in][]{crenshawetal2003}.

It is clear that some quasars affect their environment 
some of the time. The extent and the dominant mode of these
interactions remain open to interpretation. In particular, it is not clear
whether quasars are effectively removing the interstellar medium
(ISM) of their host galaxies during the high accretion rate episodes
-- those that account for the majority of the BH growth. Such feedback
has been postulated by numerical simulations
\citep[e.g.,][]{springeletal2005}, but direct observational evidence
for this process is lacking. 

In this work, we look for direct evidence of extended warm gas in
emission, using the narrow-line region (NLR) and specifically the
strong and ubiquitous \oiii$~\lambda 5007$ line.  The NLR is in some
respects the ideal tracer of the interface between the galaxy and the
active galactic nucleus (AGN), as the gas is excited by the AGN but
extended on galaxy-wide scales.  For a long time, following the
seminal work of \citet{stocktonmackenty1987}, it was thought that
truly extended emission-line regions (so-called EELRs with radii of
10-50 kpc) were only found in radio-loud objects.  Using narrow-band
imaging, these authors examined known luminous, $z \approx 0.5$ AGNs
and found that $\sim 1/3$ of the radio-loud objects had luminous extended
\oiii\ nebulosities, while none of the radio-quiet objects did.  It is
not clear if the extended gas has an internal or external origin
nor whether it is only present in radio-loud systems or is only
well-illuminated in the presence of radio jets
\citep[e.g.,][]{stocktonetal2006,fustockton2009}. 

Emission-line regions around radio-quiet systems
\citep{husemannetal2008} are not usually as extended nor as luminous
as those seen in the presence of powerful radio jets.  This
statement depends on the flux limit.  At very low surface-brightness levels ($\sim
10^{-18}$erg~s$^{-1}$~cm$^{-1}$~arcsec$^{-2}$), diverse morphologies
are observed in emission line gas \citep[e.g.,][]{colbertetal1996,veilleuxetal2003}.
An interesting exception may be the broad-line active galaxy Mrk 231.
This galaxy shows outflowing neutral and ionized gas that is extended
on $\sim 10$ kpc scales and moving at thousands of \kms\
\citep[][]{hamiltonkeel1987}.  There is a jet in this
galaxy (as well as a starburst) but the jet is not likely the source
of acceleration of the neutral outflow \citep{rupkeetal2005}.

\vbox{ 
\hskip 0.in
\psfig{file=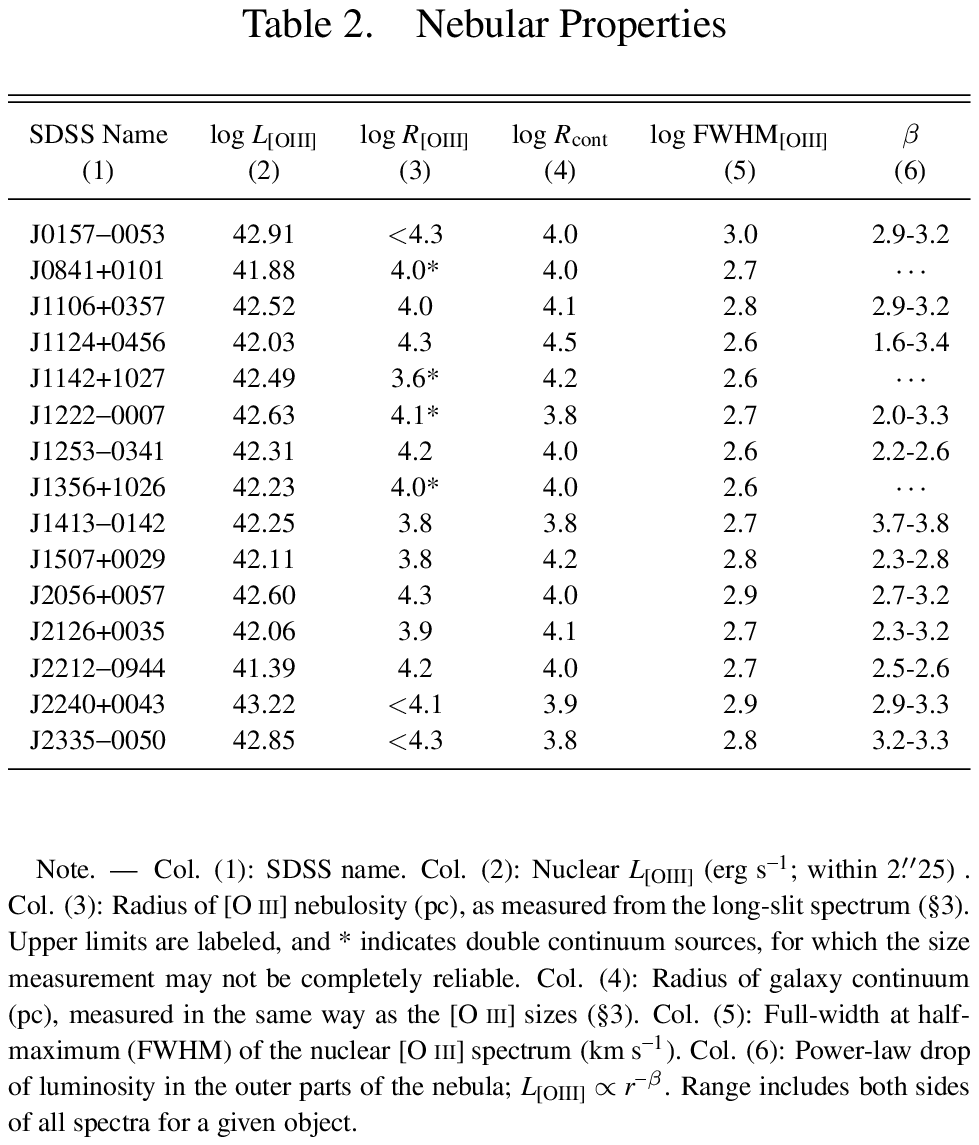,width=0.48\textwidth,keepaspectratio=true,angle=0}
}
\vskip 4mm

Rather than focus on unobscured (broad-line) quasars, where detailed
study of the NLR extent and kinematics is hampered by the presence of
a luminous nucleus, we look instead at obscured quasars.  The
experiment is worth revisiting in light of the discovery of a large
sample of obscured quasars with the Sloan Digital Sky Survey
\citep[SDSS;][]{yorketal2000}.  The sample, with $z < 0.8$, was
selected based on the \oiii\ line luminosity \citep{zakamskaetal2003}
and now comprises nearly 1000 objects \citep{reyesetal2008}.
Extensive follow-up with the \emph{Hubble Space Telescope}
\citep{zakamskaetal2006}, \chandra\ and \xmm\
\citep{ptaketal2006,vignalietal2010}, \spitzer\
\citep{zakamskaetal2008}, spectropolarimetry \citep{zakamskaetal2005},
Gemini \citep{liuetal2009}, and the VLA
\citep{zakamskaetal2004,lalho2010} yield a broad view of the
properties of the optically selected obscured quasar population.  We
target the low-redshift end of the sample, to maximize our spatial
resolution of the NLR. In our first paper, we examined the host
galaxies of our targets \citep[][Paper I hereafter]{greeneetal2009}.
Here we study the spatial distribution and kinematics of the ionized
gas.  After describing the sample and observations (\S 2), we turn to
the NLR sizes (\S 3) and then the spatially resolved kinematics of the
sample as a whole (\S 4).  We present two candidate dual obscured AGNs
(\S 5) and then summarize and conclude (\S 6).

\begin{figure*}
\vbox{ 
\vskip -0.truein
\hskip -0.1in
\psfig{file=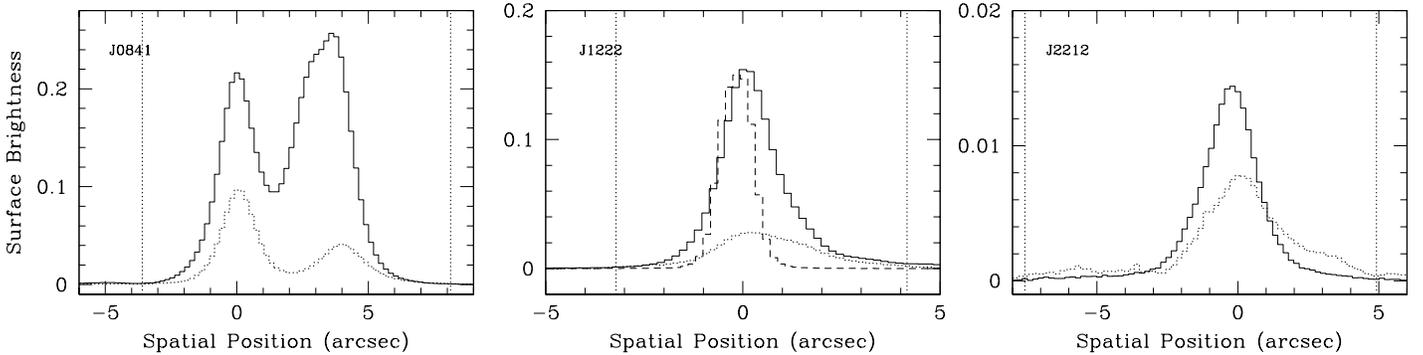,width=0.25\textwidth,keepaspectratio=true,angle=-90}
}
\vskip -0mm
\figcaption[]{
Three example [O {\tiny III}] ({\it solid}) and continuum ({\it dotted}) surface brightness
profiles as a function of spatial position, plotted in units of 
$10^{-15}$~erg~s$^{-1}$~cm$^{-2}$~arcsec$^{-2}$. 
In each case the line spectrum was summed over a velocity twice that 
of the FWHM of [O {\tiny III}], while the continuum was summed over a line-free band 
10 times wider.  The inferred [O {\tiny III}] ({\it short-dashed line}) 
diameters are shown for reference.  In the center panel, we include a flux calibration 
star for reference ({\it dashed}).
\label{fig:spat}}
\end{figure*}

\section{The Sample, Observations, and Data Reduction}

The sample and data reduction were introduced in detail in Paper I
(Table 1). The sample was selected from \citet{reyesetal2008}.  We
focused on targets with $z<0.45$ to ensure that \oiii$~\lambda 5007$ was
accessible in the observing window and imposed a luminosity cut on the
\oiii\ line of $\loiii \geq 10^{42}$~erg~s$^{-1}$ to pre-select
luminous quasars (estimated intrinsic luminosity $M_B<-24$ mag). 
Radio flux densities at 1.4 GHz were
obtained from the Faint Images of the Radio Sky at Twenty cm survey
(FIRST; \citealt{beckeretal1995}) and the NRAO VLA Sky Survey (NVSS;
\citealt{condonetal1998}). With one exception, all objects are 
radio-quiet, as determined by their position on the $\loiii - \nu
L_{\nu}$ (1.4 GHz) diagram \citep{xuetal1999,zakamskaetal2004}, and they
are at least an order of magnitude below the nominal radio-loud vs.
radio-quiet separation line in this plane. The single radio-loud
object in the sample, SDSS J1124+0456, is a double-lobed radio galaxy
(alternate name 4C+05.50) with $\nu L_{\nu}$ (1.4 GHz)$=2.9\times 
10^{42}$ erg~s$^{-1}$ which was observed with a slit nearly
perpendicular to the orientation of its large-scale radio lobes.  

We observed 15 objects over two observing runs using the 
Low-Dispersion Survey Spectrograph \citep[LDSS3;
][]{allington-smithetal1994} with a $1\arcsec \times 4\arcmin$ slit at
the Magellan/Clay telescope on Las Campanas. 
The seeing was typically $\sim 1\arcsec$ over the two runs.  We
integrated for at least one hour per target and covered one or two
slit positions (Table 1).  Lower-$z$ targets were observed with the
VPH-Blue grism in the reddest setting, for a wavelength coverage of
$4300-7050$ \AA, while the higher-$z$ targets were observed with the
bluest setting of the VPH-Red grism ($5800-9400$ \AA).  The velocity
resolution in each setting is $\sigma_{\rm inst} \approx 67$~\kms.  In
addition to the primary science targets, at least two flux calibrator
stars were observed per night and a library of velocity template stars
consisting of F--M giants was observed over the course of the run.

Since we have only long-slit observations, we do not sample the full
velocity field of the gas or stars in the galaxy.  With a few
exceptions, the galaxy images were only marginally resolved in the
SDSS images.  Thus in selecting position angles to observe we were
mainly guided by visual inspection of the color composite images.
Since these galaxies typically have very high equivalent width \oiii\
lines, we attempted to identify \oiii\ structures based on
color-gradients in the images.   As a result, the slit is not necessarily
oriented along the major or minor axis of a given galaxy.  In
particular, it is important to keep in mind when judging the radial
velocity curves of the spiral galaxies (SDSS~J1106+0357,
SDSS~J1222$-$0007, SDSS~J1253$-$0341, SDSS J2126+0035 and likely
SDSS~J1124+0456). Of these, SDSS~J1106+0357 and SDSS J2126+0035 were
observed along the major axis, and SDSS~J1222$-$0007 is within $20
\degr$ of the major axis.  The others are observed at $\sim 40 \degr$
from the major axis.  None were observed solely along the minor axis.

Cosmic-ray removal was performed using the spectroscopic version of
LACosmic \citep{vandokkum2001}, and bias subtraction, flat-field
correction, wavelength calibration, pattern-noise removal (see Paper
I), and rectification were performed using the Carnegie Observatories
reduction package
COSMOS\footnote{http://obs.carnegiescience.edu/Code/cosmos}.  For the
two-dimensional analysis discussed in this paper (e.g., the \oiii\
size determinations) we additionally use the sky subtraction provided
by COSMOS.  The flux calibration correction is determined from the
extracted standard star using IDL routines following methods described
in \citet{mathesonetal2008} and then applied in two dimensions.  In
the first paper we demonstrate that the absolute normalization of the 
flux calibration is reliable 
at the $\sim 40\%$ level.  ``Nuclear'' measurements refer to the
2\farcs25 spatial extraction.

\begin{figure*}
\vbox{ 
\vskip +0.1truein
\hskip -0.2in
\psfig{file=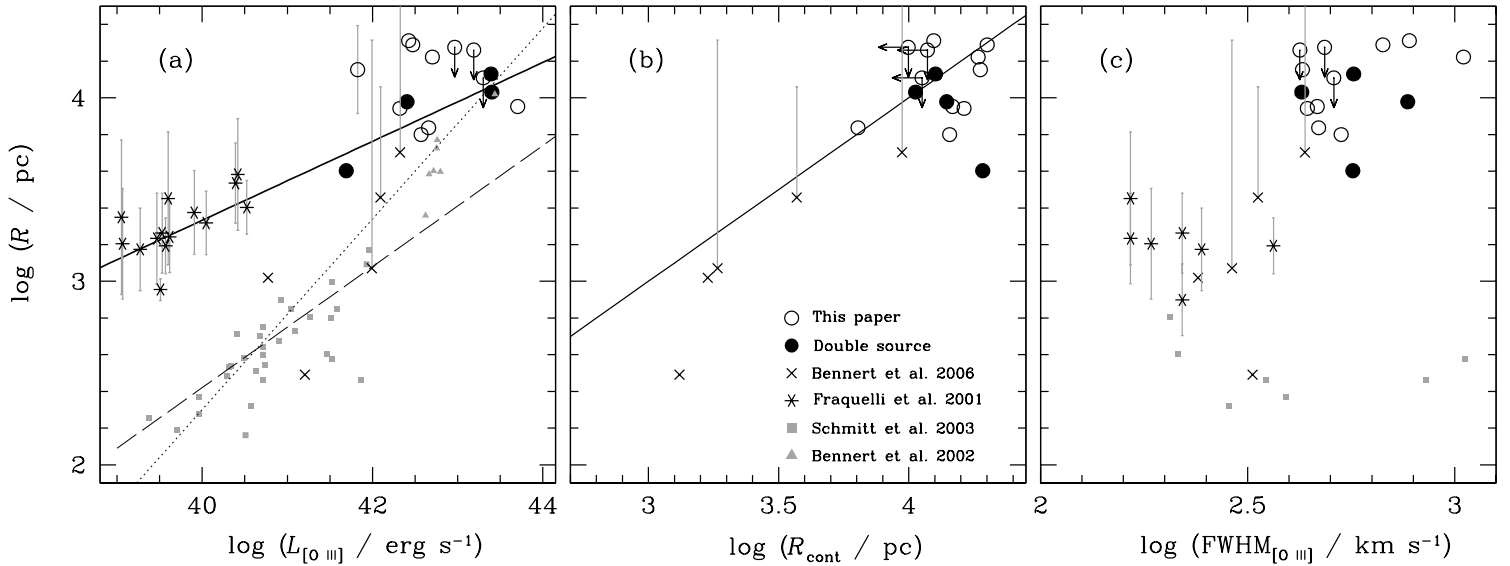,width=0.4\textwidth,
keepaspectratio=true,angle=-90}
}
\vskip -0mm
\figcaption[]{
Radii of the [O {\tiny III}]-emitting regions, as measured from
spatial cuts through the spectra centered on the [O {\tiny III}] line,
compared with various other properties. Our measurements are shown
in large circles and upper limits on the nebular sizes, in
unresolved cases, are shown as arrows.  Our estimated uncertainty of 
$\sim 3$ is demonstrated for one point only. We also show measurements
from \citet[][broad-line AGNs; {\it filled triangles}]{bennertetal2002},
\citet[][{\it grey filled squares}]{schmittetal2003}, 
\citet[][{\it asterisks}]{fraquellietal2003}, and \citet[][{\it crosses}]{bennertetal2006}.
({\it a}): $R_{\rm [O \tiny III]}$ compared to total \loiii.  The two quantities are 
correlated  (Kendall's $\tau=0.89$ with probability $P<10^{-5}$ that no correlation 
is present).  Our best-fit is shown with a solid line. Our
emission-line regions are considerably larger at a given \loiii\ than
either the broad-line quasars considered by Bennert \etal\ (2002;
  filled triangles and dotted line) or the relation derived by
Schmitt \etal\ (2003; dashed line), likely due to differences in depth. 
Thus, we exclude the latter two in our fitting.
({\it b}): $R_{\rm [O \tiny III]}$ compared to the size of the galaxy
continuum measured in a comparable fashion (see text).  Here we
include only the sample of Bennert et al. (2006) for comparison, since we do 
not have galaxy sizes available in the other cases.  In
general it is clear that the continuum and emission line sizes are of
the same order (Kendall's $\tau = 0.52$ with $P=0.098$).  
The solid line shows the 1:1 relation to guide the eye.  
({\it c}): $R_{\rm [O \tiny  III]}$ plotted against the FWHM of the [O {\tiny III}] line from the
nuclear extraction.  Here we include FWHM measurements from
\citet{whittle1992} for the Bennert et al. (2006) and Schmitt et al.
(2003) samples.  There is clearly a trend (Kendall's $\tau = 0.75, P=3 \times 10^{-4}$).
The two outliers (bottom, right) are Mrk 3 and NGC 1068.  However, we note that the 
size of NGC 1068 is likely underestimated here; \citet{veilleuxetal2003} find a size 
of $\sim 11$ kpc using a tunable filter.
\label{fig:sizes}}
\end{figure*}

\section{Nebular Sizes}

\subsection{Measurements}

The physical extent of the NLR provides one basic probe of the impact
of the AGN on the surrounding galaxy. We work with the rectified
two-dimensional spectra. In order to boost the signal in the spatial
direction, we collapse each spectrum in the velocity direction.  We
use a band with a velocity width that is twice the full width at
half-maximum (FWHM) of the nuclear \oiii\ and centered on the nuclear
\oiii\ line (Fig. \ref{fig:spat}).  The line width is measured from a
continuum-subtracted spectrum, but we do not perform continuum
subtraction on the two-dimensional spectra. This high signal-to-noise
(S/N) spatial cut allows us to measure the NLR sizes much more
sensitively than from typical narrow-band imaging.  Specifically, we
measure the total spatial extent of the line emission down to a
$5~\Sigma$ limit, where $\Sigma$ is determined from spatially-offset
regions of the collapsed surface brightness profile.  We are reaching
typical depths of $\Sigma \approx
10^{-16}$~erg~s$^{-1}$~cm$^{-2}$~arcsec$^{-2}$.  In three cases the
nebular spectra are not spatially resolved (i.e., the spatial
distribution matches that of a standard star). There are six objects
for which we have multiple slit positions. The range in nebular size
derived from cases with multiple slit positions is $\sim 30\%$.  

In a few cases (SDSS~J1356$-$1026, SDSS~J2126$+0035$ \&
SDSS~J2212$-0944$), the line ratios change as a function of radius and
\oiii/\hbeta\ falls below three.  This changing ratio may reflect
changes in the ionization parameter or gas-phase metallicity, or a
transition from ionization dominated by the AGN to \ion{H}{2} regions
\citep[e.g.,][]{bennertetal2006}. By ionization parameter, we mean 
the ratio of the density of ionizing photons to the density of electrons.
Given the luminosities of quasars in
our sample and the rates of star formation in their hosts
\citep{zakamskaetal2008}, we expect that the number of ionizing
photons from the quasars exceeds that from stars by about an order of
magnitude. Nevertheless, since quasar illumination is not necessarily
isotropic and since photons from star formation are distributed more
uniformly within the galaxy than those arising from the central
engine, it is plausible that we may see gas excited by stars in the
outer regions of the galaxy. Shock excitation is unlikely since the
linewidths are uniformly narrow in these outer regions.  To be safe,
we exclude the regions with \oiii/\hbeta$<3$ when calculating the NLR
sizes. We have not applied any correction for reddening, which could
be substantial
\citep[e.g.,][]{melendezetal2008red}. \citet{reyesetal2008} show that
deriving robust extinction corrections for the SDSS obscured quasars
is not straightforward, and we neglect such corrections here.

One of the objects in our sample, SDSS J1356+1026, has a much more
dramatic extended emission-line nebula than the rest (Figure
\ref{fig:bubbleim}).  We will discuss the detailed kinematics and
energetics of this object in more detail in a parallel paper (Greene
\& Zakamska in preparation).  For the present work, we explore the
implications of detecting one single extended emission-line region in
the sample (\S 6).

We should note that deriving nebular sizes is an ill-defined task.
First of all, ionized nebulae need not have regular shapes, and so the
definition of size is not necessarily well-defined.  This difficulty
is only exacerbated when long-slit spectra are used to define the
size, since our slit may well miss spatially extended regions.
Furthermore, the concept of size depends sensitively on the depth
of the observation.  Deep observations probing depths of a few times
$10^{-18}$~erg~s$^{-1}$~cm$^{-2}$~arcsec$^{-2}$ indeed reveal faint,
extended gas with a range of morphologies
\citep[e.g.,][]{veilleuxetal2003}.  Thus, the primary size
uncertainties are in these systematics, which dwarf the measurement
errors.  

We quantify the uncertainties in the following manner.  First of all,
since we have not included surface brightness dimming, there is a
dispersion of $\sim 0.1$ dex in the sizes due to distance.  Secondly,
and more important, narrow-line regions are not strictly round.  Thus,
depending on the position of the long-slit, we may derive a different
answer.  We have found, for the six objects with multiple slit
positions, that the sizes agree within $30\%$.  Finally, and most
difficult to quantify, the shape will likely grow more irregular as we
push to lower flux limits.  We have attempted to quantify this
dispersion in emission line profile using the ratio between the
luminosity-weighted mean width of each spatial \oiii\ profile and the
adopted radius measured down to a fixed surface brightness.  Were the
NLRs all of the same shape, then the mean width would be a fixed fraction of the 
total size.  Instead, the ratio ranges from 0.2 to 4, with a typical value of three.
We thus adopt a factor of three as the
overall uncertainty in the sizes. Additionally, we flag as
particularly uncertain those systems with a nearby massive companion
galaxy, since there we are further contaminated by tidal gas.

\subsection{Obscuration, Ionization, and Excitation}

NLR sizes have been measured from narrow-band imaging
\citep{mulchaeyetal1996im,bennertetal2002,schmittetal2003} and from
long-slit spectroscopy
\citep[e.g.,][]{ungeretal1987,fraquellietal2003,bennertetal2006}.
Narrow-band imaging is preferable for studying the NLR morphology, but
reaches shallower limits than the spectroscopy.  

\vbox{ 
\vskip -0.truein
\hskip -0.1in
\psfig{file=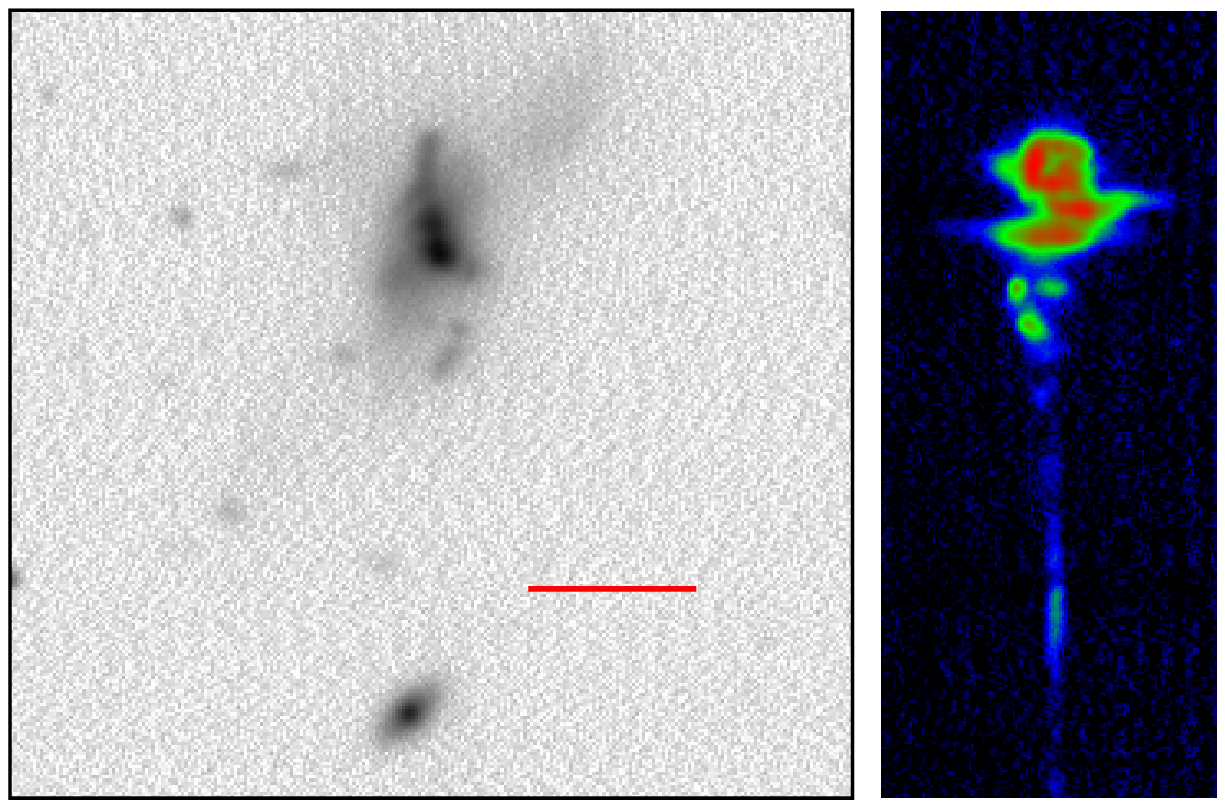,width=0.3\textwidth,keepaspectratio=true,angle=-90}
}
\vskip -0mm
\figcaption[]{
{\it Left}: Image ($r-$band) of the merging galaxies SDSS J1356+1026. Scale 
bar indicates 10\arcsec\ (22 kpc).  N is down and E to the right.  
{\it Right}: The two-dimensional spectrum centered on the continuum-subtracted 
[O {\tiny III}]$~\lambda 5007$
line.  Oriented as the image.  The spectrum spans 1814 \kms in the velocity (x) 
direction.
\label{fig:bubbleim}}
\vskip 5mm
\noindent
Integral-field
observations allow one to study two-dimensional kinematics
\citep{humphreyetal2010}, but for local objects only cover the inner
NLR \citep[e.g.,][]{barbosaetal2009}.

We compile a comparison sample of lower-luminosity obscured AGNs with
measured NLR sizes from the literature
\citep{bennertetal2002,schmittetal2003,bennertetal2006,
  fraquellietal2003}.  We include the Bennert et al. (2002) and
Schmitt et al. (2003) measurements in Fig. 2 for completeness, but note that the
sizes cannot be compared directly with those we measure here, because
of the difference in depth.  The limiting surface brightness values
that we achieve in this work are at least a factor of 10 deeper than
these narrow-band imaging studies from space, which range from $\sim
10^{-15}-3 \times 10^{-14}$~erg~s$^{-1}$~cm$^{-2}$~arcsec$^{-2}$.  For
this reason, we do not include the space-based measurements in any
analysis presented here (e.g., fitting of relationships).  Fraquelli
et al. do not quote sizes but rather provide power-law fits to the
surface brightness as a function of distance to the nucleus. Taking
their functional form, we calculate sizes that match our limiting
surface brightness of $\sim
10^{-16}$~erg~s$^{-1}$~cm$^{-2}$~arcsec$^{-2}$.  For uniformity, we
calculate sizes for Bennert et al. (2006) in the same way, and we
adopt their smaller radii in cases where star formation dominates in
the outer parts.  In cases of overlap between works, we prefer the
\citet{bennertetal2006} observations, since they are both sensitive
and take into account photoionization by starlight.  The measurements
for our sample are summarized in Figure~\ref{fig:sizes} and Table 2,
while the comparison samples are shown in Figure~\ref{fig:sizes}.

The observed distribution of NLR gas depends on the geometry and
luminosity of the ionizing source, the geometry and kinematics of the
host ISM (e.g., disk, spherical, outflow, or infall), and the density
distribution of the gas. While most of these are likely related to the
morphology and dynamical state of the galaxy, the geometry of the
ionizing source is tied to the
orientation of the AGN. In the simplest model, the galaxy ISM is
spherically distributed, while the ionizing radiation from the AGN
emerges anisotropically along lines of sight unaffected by the
circumnuclear `torus', as postulated by unified models of AGN
activity. In this case we expect to see ionization cones when the beam
is not pointed directly at us, reflecting the geometry of
circumnuclear obscuration. Such cones are observed in images of nearby
Seyfert galaxies
\citep{pogge1988,tadhuntertsvetanov1989,storchibergmannetal1992} and
more recently in the luminous obscured quasars studied here
\citep{zakamskaetal2006}.

\begin{figure*}
\vbox{ 
\vskip -0.5truein
\hskip 0.15in
\psfig{file=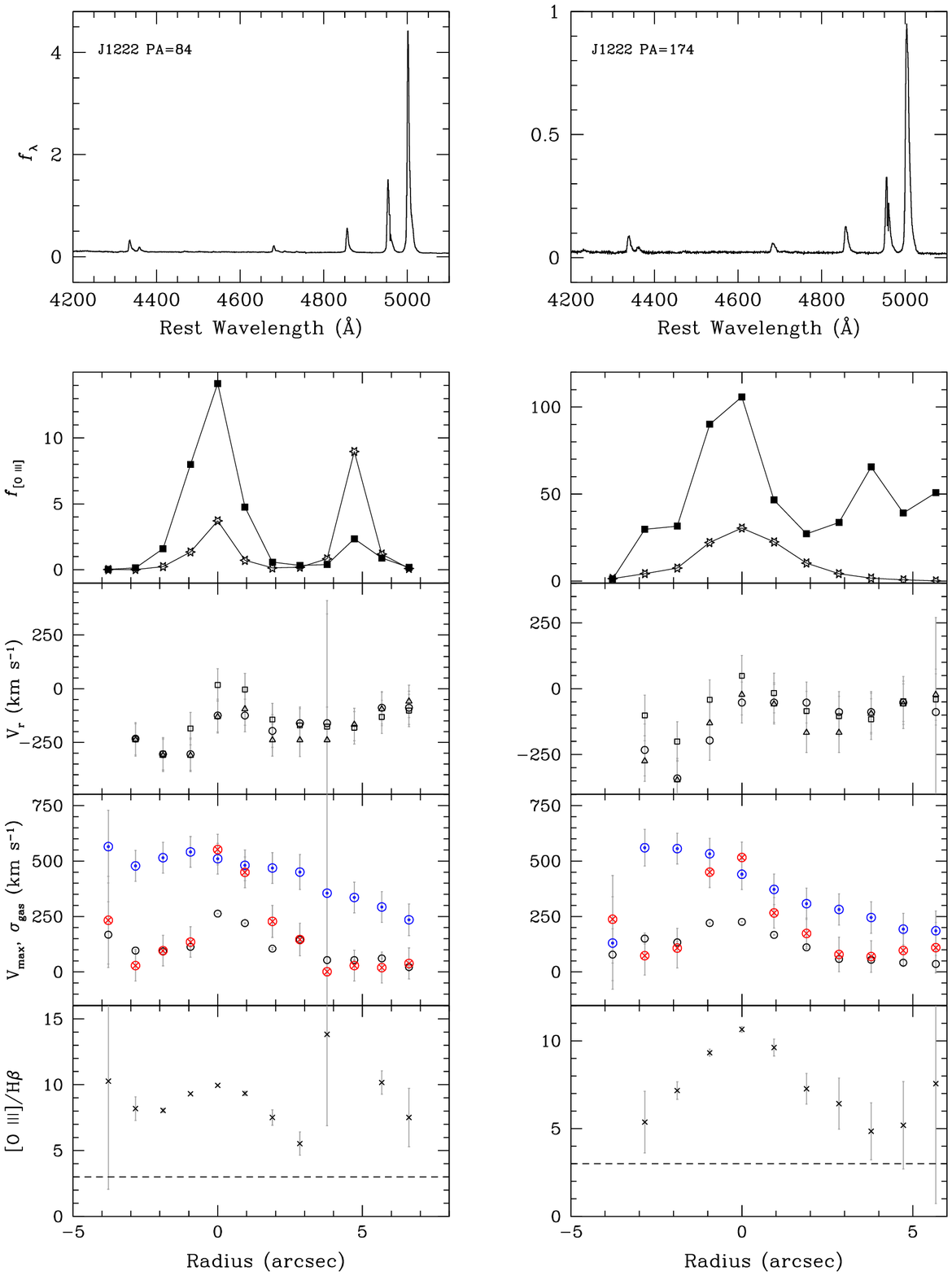,width=0.8\textwidth,keepaspectratio=true,angle=0}
}
\vskip -0mm
\figcaption[]{
\small
Resolved spectroscopic measurements for SDSS~J1222$-0007$.  
In the top panel we show the 
central spectrum, extracted within the inner $\sim 1\arcsec$, arbitrarily 
normalized.  Then we 
show the flux in the [O {\tiny III}] line (filled black squares) and 
the continuum (open stars) as a function of radius 
(10$^{-15}$~erg~cm$^{-2}$~s$^{-1}$).  
In the third panel we show the radial velocity curve of the peak of 
[O {\tiny III}] ({\it open triangles}), the velocity of the peak of the H$\beta$ line 
({\it open circles}), and the luminosity-weighted mean 
velocity of [O {\tiny III}] ({\it open squares}).  In the fourth panel the velocity 
dispersion of the [O {\tiny III}] line ({\it open black circles}) is compared with 
the maximum velocity (the velocity at $20\%$ of the line maximum) 
to the red ({\it red open circle with cross}) and the blue 
({\it blue open circle with dot}). In the final panel the ratio of 
[O {\tiny III}]/\hbeta\ is plotted ({\it black crosses}). A ratio of [O {\tiny III}]/\hbeta\ 
of three is noted with the dashed line.
\label{fig:rot1222}}
\end{figure*}

In this simplest geometry, we would expect to find smaller sizes in
unobscured sources, when looking closer to the axis of the ionization
cones.  The difference in distributions depends on the expected
opening angle of the torus, with larger difference for smaller opening
angles.  Recent observations of obscured quasars suggest that the
space densities of obscured and unobscured sources are $\sim$ equal
\citep{reyesetal2008}, leading to opening angles of $\lesssim 120
\degr$, but even if significantly smaller opening angles are assumed,
the expected differences in the median projected size between the two
populations is small ($\la 0.2$ dex). At low redshift and (thus) lower
luminosity, ionization cones are also observed in unobscured sources
\citep[e.g.,][]{evansetal1993,boksenbergetal1995}, while round NLRs
are observed in both types \citep{mulchaeyetal1996}.  Presumably, the
ISM is not always spherically distributed or relaxed
\citep{mulchaeyetal1996,schmidtetal2007}.

In Figure~\ref{fig:sizes}{\it a} it appears that the NLR sizes of our
obscured quasars are larger than the unobscured ones (median
difference 0.4 dex).  However, this difference can be explained by
differences in the depths of the observations, thus we cannot address
orientation differences in detail from these samples.  It is
interesting to note that at lower luminosities,
\citet{schmittetal2003} do not see a significant size difference
between the two populations.  These observations, while shallow, are
uniform between the obscured and unobscured populations.

There has been some debate in the literature about the slope of a
purported correlation between the NLR size and the AGN luminosity
\citep[e.g.,][]{bennertetal2002,schmittetal2003,bennertetal2006}. Some
correlation is expected, given that the AGN is photoionizing the NLR
gas, but the form it takes may tell us something about covering factor
or density as a function of luminosity.  It is clear from Figure
\ref{fig:sizes}({\it a}) that generally larger NLRs are found in more
luminous objects (Kendall's $\tau=0.89$ with probability $P<10^{-5}$
that no correlation is present).  It is also clear that there is
substantial scatter; we find an rms scatter of 0.3 dex in radius at
fixed \loiii.  We performed Monte Carlo simulations of ionization
cones observed at random directions (restricted to be outside the
cones).  They suggest that the orientation of the NLR axis relative to
the line of sight is not a significant source of the observed
scatter. At a fixed NLR size, orientation effects introduce a scatter
of $<0.15$ dex within each (obscured or unobscured) subpopulation,
even when a wide range of opening angles is allowed for. Therefore, the
observed scatter is likely due to the combination of the true variance in NLR
sizes at a given luminosity and to the differences in the definition
of NLR ``size''.  For instance, \citet{bennertetal2006} derive sizes
that are factors of $\sim 2$ larger than those based on \hst\
narrow-band imaging because of their increased sensitivity.  Given
that the NLR is not always spherically symmetric or smooth, defining a
meaningful size that is insensitive to depth is a difficult problem.

For completeness, we fit a power-law relation between \loiii\ and NLR
size, using all narrow-line comparison samples as well as the objects
considered here.  Because there are upper limits on the sizes, we
calculate a linear regression using the binned Schmitt method, from
the Astronomy Survival Analysis (ASURV) software as implemented in
{\tt iraf} \citep{feigelsonnelson1985,isobeetal1986}.  The fit is
shown in Figure 2.  We find:
\begin{eqnarray}
{\rm log} (R_{\rm NLR} / 10^3 {\rm pc})  = 
(0.22 \pm 0.04) \,{\rm log} (L_{\rm [O {\tiny III} ]} / 10^{42} {\rm erg~s^{-1}})  \\
+(3.76 \pm 0.07). \nonumber
\end{eqnarray}

The shallow slope we observe is consistent with a picture in which the
nebulae are matter-bounded. At the distances from the
quasar that we are probing with our observations, the density of
material is low enough that the emissivity is no longer limited by the
flux of photons by the quasar, but rather by the low density of the
gas, and a large fraction of photons can escape into the intergalactic
medium.  Note that the correlation between AGN continuum luminosity
and \loiii\ in broad-line AGNs \citep[e.g.,][]{yee1980} suggests that
the nebulae are limited by the number of photons in the bright central
regions of the galaxy, but that the situation changes in the diffuse
outer parts.  If so, we would expect size to scale as the square-root
of luminosity at low luminosities and then flatten out to at high
luminosities, modulo differences in host galaxies.

In addition to measuring the nebular sizes, we also parameterize the 
luminosity drop in the outer parts as a power-law and measure the 
power law slope (\loiii$\propto r^{-\beta}$).  The slopes range from 
$1.6 < \beta < 3.8$ (Table 2).  These slopes correspond to density profiles 
with slopes ranging from 1.3 to 2.4, in good agreement with the \hst\ 
observations of \citet{zakamskaetal2006}.

One concern, as pointed out by \citet{netzeretal2004}, is that
eventually the ionizing photons will run out of interstellar medium to
ionize, particularly in the most luminous quasars.  The NLR size
cannot in general grow indefinitely beyond the confines of the host
galaxies.  In Figure \ref{fig:sizes}({\it b}), we compare the
continuum and nebular sizes.  Rather than using effective radii of
host galaxies from photometry, we use the same method to measure the
continuum extent as we used for the \oiii\ lines, collapsing the
two-dimensional spectrum in the spectral direction over line-free
regions to boost the signal.  Galaxy sizes are weakly correlated with
NLR size (Kendall's $\tau = 0.52$ with $P=0.0975$). We see that the
NLR sizes are comparable to the galaxy continua.  The exception is
SDSS J1356+1026, which contains the spectacular bubble shown below.
At these luminosities, the AGN is effectively capable of photoionizing
the entire galaxy ISM, as well as companion galaxies out to several
tens of kpc, as we saw in some of our previous long-slit observations
\citep{liuetal2009}.

Outflowing components of the NLR are routinely seen in radio galaxies
\citep[e.g.,][]{mccarthy1993,villarmartinetal1999}, as well as in
Seyfert galaxies \citep[e.g.,][]{crenshawkraemer2000,rupkeetal2005}.
On small scales, detailed modeling of the inner ($<300$ pc) NLRs of a
few local AGNs with \hst\ indicates a surprising uniformity in
behavior, with $v \propto r$ along an evacuated bicone
\citep{crenshawkraemer2000,crenshawetal2000,ruizetal2001}.
Interestingly, we see similar qualitative behavior in SDSS J1356+1026
(Greene \& Zakamska in prep).  However, in general, NLR kinematics on
larger scales are not as uniform, with mechanisms ranging from jet
acceleration to radiation pressure driving
\citep[e.g.,][]{ruizetal2001,grovesetal2004, rupkeetal2005}.  For a
complete review, see \citet{veilleuxetal2005}.  We do find some correlation 
between FWHM and luminosity 
(Figure \ref{fig:sizes}{\it c}; Kendall's $\tau = 0.75, P=3 \times 10^{-4}$). 
We will argue below based on the observed large velocity dispersions 
at large radius that the AGN energy is stirring up the gas on large scales, thus 
explaining this correlation.

\section{Two-Dimensional Analysis}

\vbox{ 
\vskip +0.3truein
\hskip -0.3in
\psfig{file=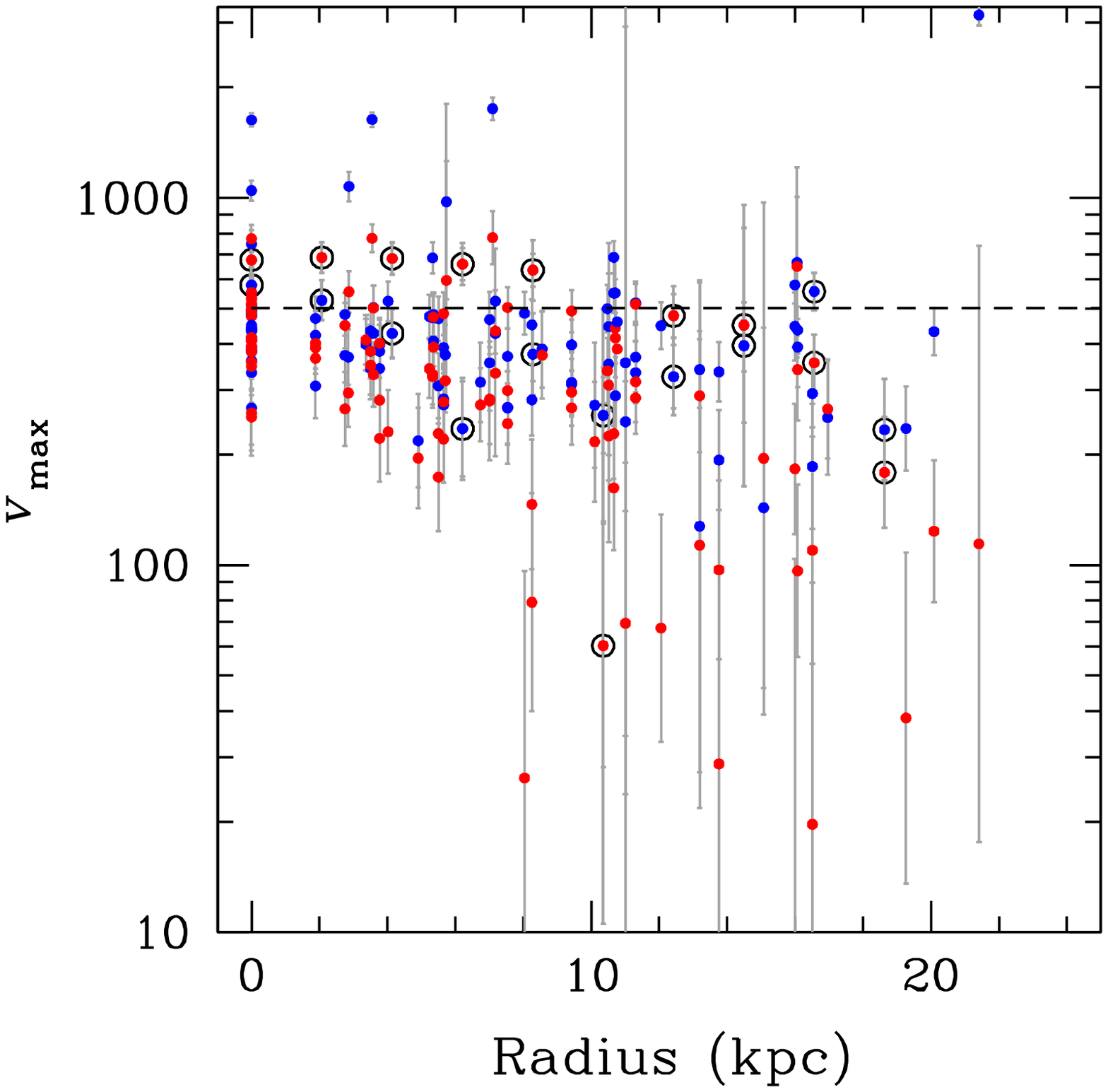,width=0.5\textwidth,keepaspectratio=true,angle=0}
}
\vskip -0mm
\figcaption[]{
Maximum velocity as a function of radius.  The maximum velocity is measured 
as the velocity at 20\% of the line peak, as referenced to the systemic 
velocity of the stars.
At each position we show the maximum red- (red circle) and blue-shifted (blue circle) 
velocity.  While
there is extended emission in nearly all targets, the typical
velocities are not higher than $\sim 400$~\kms\ for the majority of
the targets.  SDSS J1356+1026 is highlighted with large black circles. 
The dashed line highlights a velocity of $500$~\kms, which is the approximate 
escape velocity for these galaxies at $r_e$.  We note that the effective radii of these
galaxies, for which we have well-resolved imaging, range from $1 \pm
0.4$ to $11 \pm 2$ kpc, with a median of $3 \pm 0.6$ kpc.
\label{fig:maxvel}}
\vskip 5mm

In this section we present the results of our two-dimensional analysis
on the long-slit spectra.  First we present velocity and dispersion
profiles, as well as emission line ratios, as a function of
position. To obtain these measurements, we extract spectra at uniform
intervals as a function of spatial position along the slit.  We start
with rectified two-dimensional spectra from COSMOS.  Each spectrum is
extracted with a width of 0\farcs95 (5 pixels) to match the typical
seeing of the observations.  The central spatial position is
determined by the spatial peak in the \oiii\ emission.  The systemic
velocity is determined from the absorption lines.  Galaxy continuum
subtraction is performed for each spectrum using a scaled version of
our best-fit model from the nuclear spectrum, with only the overall
amplitude allowed to vary.  While this is not strictly speaking a
correct model, we have insufficient S/N in the off-nuclear spectra to
constrain velocity or velocity dispersion, let alone changes in
stellar populations.

Once the continuum-subtracted spectra are in hand, we fit the
\hbeta+\oiii$~\lambda \lambda 4959, 5007$ lines for each spectrum as
in Paper I \citep[see also][]{hfs1997broad,greeneho2005o3}.  Each line
is modeled as a sum of Gaussians (a maximum of two for \hbeta\ and
three for \oiii). The relative wavelengths of each transition and the
ratio of the \oiii\ lines are fixed to their laboratory values, but
the central velocity and line widths are allowed to vary from spectrum
to spectrum.  From these fits we are able to derive velocity, velocity
dispersion, and line-ratio profiles as a function of spatial position.
We report three measures of velocity at a given position, the peak in
the \oiii\ line, the peak in the \hbeta\ line, and the flux-weighted
mean velocity in the \oiii\ line. The velocity dispersion is measured
as the FWHM of the \oiii\ model divided by 2.35.  At each spatial
position we also measure the ``maximum'' and ``minimum'' velocities as
the velocities at $20\%$ of the \oiii\ peak intensity
\citep[e.g.,][]{rupkeetal2002} relative to the systemic velocity of
the stars (shown as blue bullseyes and red crosses, respectively, in
Figure \ref{fig:rot1222}.)

Errors are derived from Monte Carlo simulations.  For each
spectrum we generate 100 mock spectra using the best-fit parameters
at that radius and the S/N of the original spectra.  We fit each mock
spectrum and the quoted parameter errors encompass $68\%$ of the mock
fit values.

\subsection{Radial Velocity Curves}

In Figure \ref{fig:rot1222} we present a representative radial
velocity curve for SDSS~J1222$-0007$.  The remainder are shown in the
Appendix.  First, we note that overall the radial velocity curves are
flat.  In Paper I we presented detailed two-dimensional photometric
fitting of these galaxies (with the exception of SDSS~J1124+0456 and
SDSS~J1142+1027).  Using these fits, we divide the sample by the
bulge-to-total ratio (B/T), and call galaxies with B/T$\leq 0.1$
disks (SDSS~J1106+0357, SDSS~J1222$-$0007, SDSS~J1253$-$0341, and
probably SDSS~J1124+0456), while the rest are bulge dominated.
Additionally, those with clear tidal signatures are ``disturbed''
(SDSS~J0841+0101, SDSS~J1222$-$0007, SDSS~J1356+1026, and
SDSS~J2212$-$0944).  We would expect to see the signature of rotation
most clearly in disk-dominated galaxies.  We note once again that
SDSS~J1106+0357 and SDSS J2126+0035 were observed along the major
axis, SDSS~J1222$-$0007 was within $20 \degr$ of the major axis, and
the remaining two galaxies were observed at a $\sim 40\degr$ angle to
the major axis.  We would expect to see the signature of rotation 
in most of these galaxies.  Instead, we only see rotation in the 
case of SDSS~J1106+0357, SDSS~J1124+0456, SDSS~J1142+1027, and
SDSS~J2212$-$0007.  Although with such a wide range of position angles, 
and such a small sample, it is hard to say for sure, we find it suggestive that 
neither SDSS~J1253$-$0341 nor SDSS J2126+0035 shows rotation.

The sample galaxies showing rotation in their radial velocities also
tend to show declines in \sigmagas\ by factors of two or more in the
outer parts (e.g., SDSS~J1124+0456).  In contrast, those galaxies with flat
radial velocity curves (the majority in this sample) also have notably flat
\sigmagas\ distributions at kpc scales.  Again, this is strongly in
contrast to the kinematics in the stars, even in bulge-dominated systems
\citep[e.g.,][]{jorgensenetal1995}.  More to the point, it is in
contrast to the kinematics of warm gas in 
inactive late-type \citep[e.g.,][]{pizzellaetal2004} and early-type
\citep[e.g.,][]{fillmoreetal1986,bertolaetal1995,vegabeltranetal2001}
spiral galaxies.

In Paper I we showed that \sigmagas\ in the nucleus is uncorrelated
with \sigmastar.  Again, this behavior is in striking contrast not
only to inactive galaxies but also to local, lower-luminosity active
galaxies, for which it has been long been known that on average
\sigmagas/\sigmastar$\approx 1$
\citep[e.g.,][]{nelsonwhittle1996,greeneho2005o3,ho2009o3}.  Here we
are making a stronger statement.  Not only is the luminosity-weighted
gas dispersion uncorrelated with the dispersion in the stars, but the
dispersion in the gas stays high out to kpc scales in these galaxies.
These observations provide new reason to doubt that gas
velocity dispersions can be substituted for stellar velocity
dispersions in luminous AGNs \citep[e.g.,][]{shieldsetal2003,
  salvianderetal2007}.  

\vbox{ 
\vskip +0.3truein
\hskip -0.2in
\psfig{file=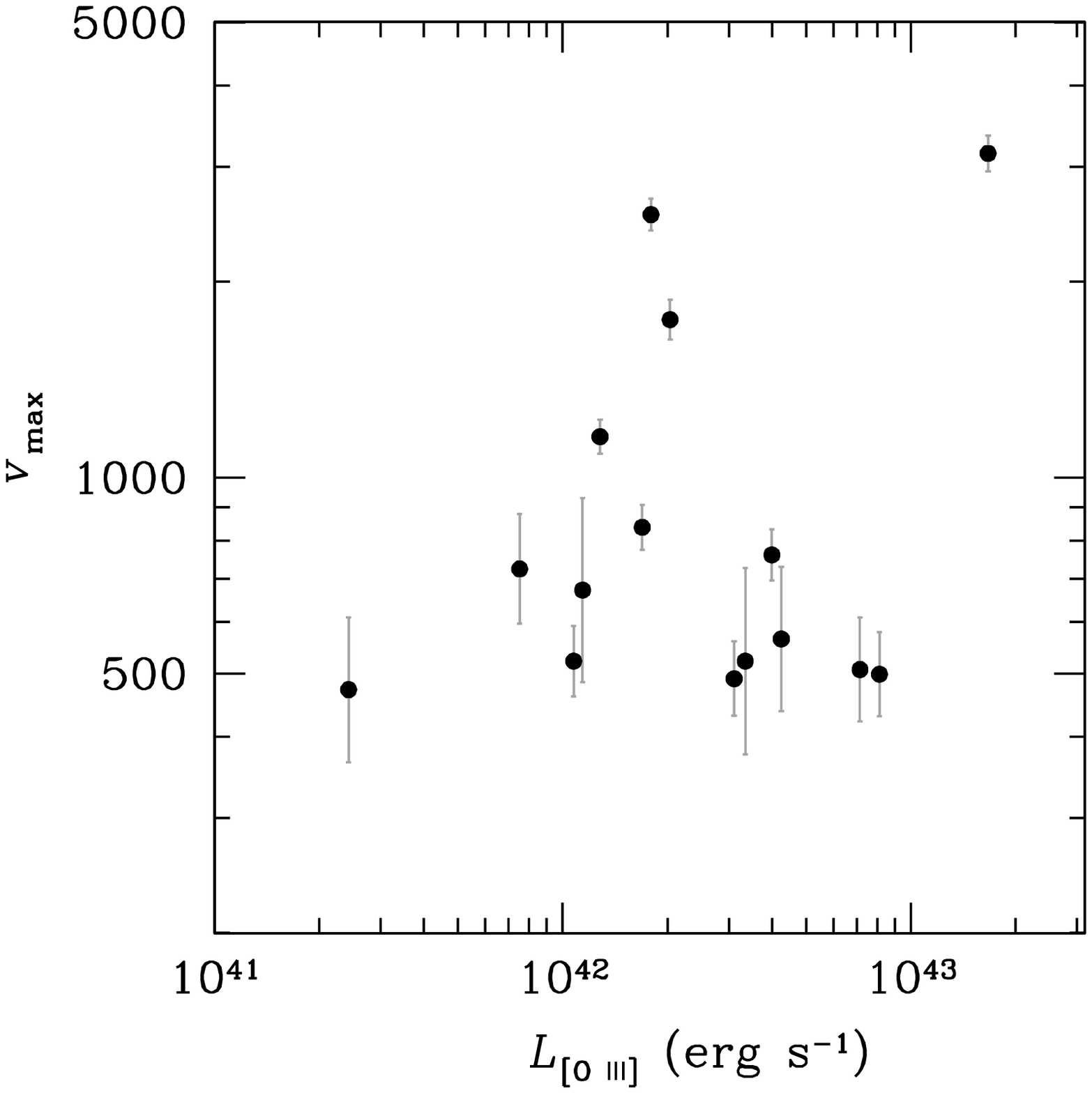,width=0.45\textwidth,keepaspectratio=true,angle=0}
}
\vskip -0mm
\figcaption[]{
Nuclear [O {\tiny III}] luminosity compared with the maximum gas velocity 
for each object.  The maximum velocity plotted here is the largest that we 
measure for any individual object, although we exclude the maximum velocity in 
the nuclear spectrum.  No correlation 
is apparent between the two (Kendall's $\tau = 0.13$ with a probability, 
$P=0.73$ of no correlation).
\label{fig:o3maxvel}}
\vskip 5mm

This behavior is different from that seen in regular inactive
galaxies.  It is also different from that in local, well-observed
Seyfert galaxies. Previous work looking at the kinematics of
lower-luminosity local Seyfert galaxies has found evidence for a
two-tiered NLR structure \citep[e.g.,][]{ungeretal1987}.  In such
objects, the inner or classical NLR extends to a few hundred pc and
has linewidths of $\gtrsim 500$~\kms.  At higher spatial resolution,
there is clear evidence for outflow in the inner hundreds of pc in
well-studied objects
\citep[e.g.,][]{crenshawetal2000,crenshawkraemer2000,ruizetal2001,
  barbosaetal2009}. In contrast, at larger radius, the linewidths drop
and the kinematics of the NLR gas simply reflect that of the bulge or
disk in which the gas sits \citep[e.g.,][]{nelsonwhittle1996,
  greeneho2005o3,walshetal2008}. Clearly, the observed kinematics of
gas in hosts of obscured quasars are quite dissimilar from this
picture.

Many of the host galaxies of our obscured quasars have nearby
companions and/or show signs of recent interactions. It is therefore
possible that the gas is being stirred by gravitational interactions
with nearby galaxies.  To explore that possibility further, we examine
the analogous inactive ultra-luminous infrared galaxies.  The
integral-field spectra of \citet{colinaetal2005} show that even in
these ongoing mergers the gas kinematics traces that of the stars.
The \sigmagas\ profile is typically seen to decline in the outer parts
as in non-merging systems, again in contrast to our findings for hosts
of obscured quasars. Of course, there are exceptions in the Colina et
al. sample, where the gas velocity dispersions are very complex.  On
the other hand, the mergers are more advanced in general than in our
sample.  Thus, while we cannot rule out gravitational effects in all
cases, it seems most likely that the nuclear activity is directly
responsible for stirring up the gas.  We now address whether there is
evidence for bulk motions (e.g., large-scale outflows) in the gas
based on the kinematics.

\vbox{ 
\vskip +0.4truein
\hskip -0.2in
\psfig{file=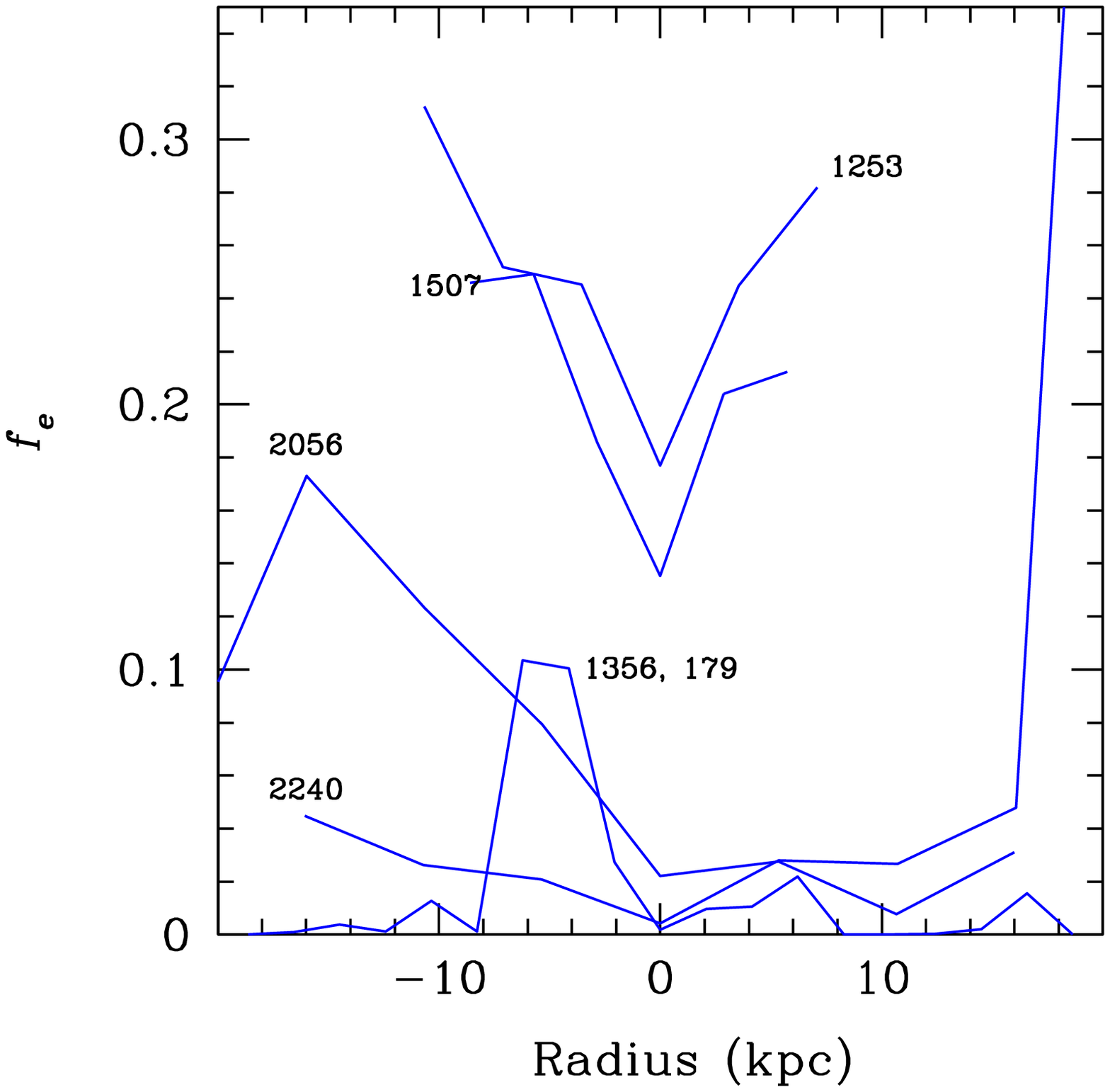,width=0.4\textwidth,keepaspectratio=true,angle=0}
}
\vskip -0mm
\figcaption[]{
Fraction of the [O{\tiny III}] emission line luminosity emerging at a velocity 
greater than the escape velocity for all objects where the escaping fraction 
exceeds 10\% in at least one shell.  Object names are labeled with the 
first four digits of the full position. The escape velocity is 
approximated from the measured velocity dispersion assuming 
that $v_{c} = \sqrt{2} \sigma_{\ast}$ and that the halo has 
a radius of 100 kpc.  We then simply integrate the part of the line with 
velocities greater than the escape velocity.  Here we show the fraction 
of the line emission that is in this high-velocity component as a function of 
radius from the slit center.  
\label{fig:fracesc}}
\vskip 5mm
\noindent

\subsection{Maximum Velocities}

We have derived ``maximum'' red- and blueshifted velocities at $20\%$
of the line profile, relative to the systemic velocity of the stars.
We examine the distribution of maximum velocities as a function of
radius for the ensemble of spectra in Figure \ref{fig:maxvel}.  While
the emission extends to kpc scales for the majority of the targets,
the gas velocities are not typically very high.  The median maximum
blue velocity at 8 kpc is $\langle v_{\rm blue} \rangle = 400 \pm
70$~\kms\ while towards the red it is $\langle v_{\rm red} \rangle =
330 \pm 50$~\kms, where we quote errors in the mean.  A few objects
(SDSS J1253$-0341$, SDSS J1222$-0007$) have gas at velocities
exceeding 500~\kms.  We note that the effective radii of these
galaxies, for which we have well-resolved imaging, range from $1 \pm
0.4$ to $11 \pm 2$ kpc, with a median of $3 \pm 0.6$ kpc. These
velocities exceed the velocity dispersions of the galaxies, but they
do not compare to the $\sim$ thousands of \kms\ outflow velocities
seen by \citet{tremontietal2007} and postulated to be driven by recent
AGN activity.  Furthermore, they are not close to the escape velocity
needed to actually unbind the gas.  As we show in Figure
\ref{fig:o3maxvel}, there is no evidence for a correlation between the
nuclear \loiii\ luminosity and the maximum observed velocity
(Kendall's $\tau = -0.10$ with a probability $P=0.8$ of no
correlation).

\begin{figure*}
\vbox{ 
\vskip +1.3truein
\hskip +0.5in
\psfig{file=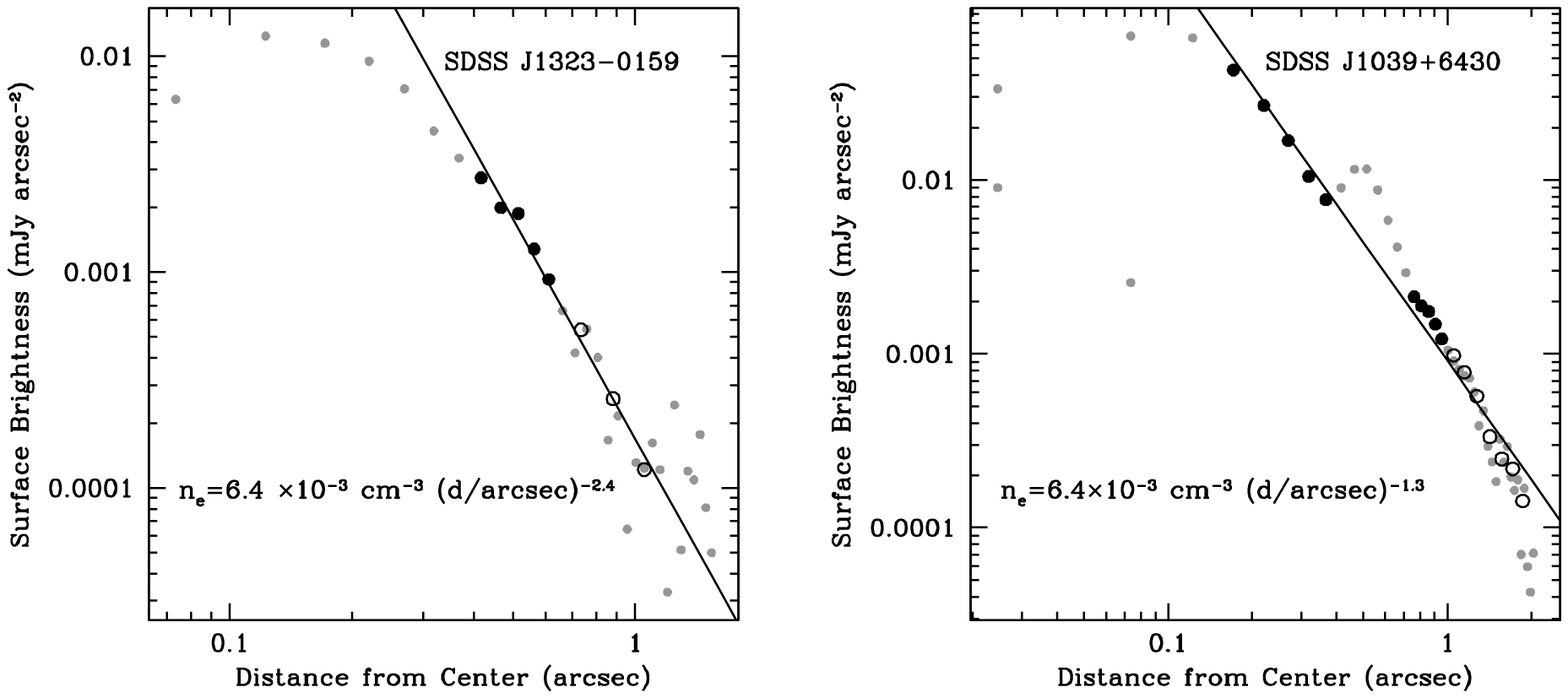,width=0.85\textwidth,keepaspectratio=true,angle=0}
}
\vskip -0mm
\figcaption[]{
The measurements of the scattered light emission made in the
rest-frame UV band using the \emph{HST} can be translated into a constraint
on the density of scattering particles. Grey points are all measurements 
from the \hst\ images of Zakamska et al. (2006), while fitting was done  
with the large symbols, and open symbols are binned. The observed surface
brightness of scattered light (vertical axis) as a function of
distance $r$ from the center of the galaxy (horizontal axis) is
proportional to the illuminating flux from the central AGN, $L/(4\pi
r^2)$, to the scattering cross-section of particles, ${\rm d}
\sigma/{\rm d} \Omega$, and to the column density of scatterers
$\simeq n_s r \beta$, where $\beta$ is the opening angle of the
scattering region. Solid lines show the best power-law fit to electron
density necessary to reproduce the scattered emission, assuming that
scattering is due to dust particles. Per given hydrogen density, dust particles are
about 60 times more efficient scatterers than electrons, even if the
gas is fully ionized. The measurements are made using
host-galaxy-subtracted images at about rest-frame 3100\AA. The deficit
of light in the central parts is due to the PSF smearing and dust
obscuration.
\label{fig:scatter}}
\end{figure*}

We now quantitatively address whether any of the gas is approaching
the escape velocity.  Following \citet[][]{rupkeetal2002}, we
calculate an approximate escape velocity for each galaxy by assuming
that the circular velocity scales with the velocity dispersion as
$v_{c} = \sqrt{2} \sigma_{\ast}$.  Assuming the potential of an isothermal sphere, the
escape velocity as a function of radius scales as:
\begin{equation}  
v_{\rm esc}(r) = \sqrt{2} v_{c} [1 + {\rm ln} (1 + r_{\rm max}/r)]^{0.5}.
\end{equation}
Although $r_{\rm max}$ is unknown, the escape velocity depends only
weakly on its value.  Thus we assume $r_{\rm max} = 100$~kpc in all
cases.  The escape velocities thus estimated range from 500 to 1000
\kms\ over the entire sample, but only vary by $\sim 60\%$ for an
individual object over the range of radii that we probe.

With escape velocities in hand, we can now address what fraction of
the line emission comes from gas that is moving at or above the escape
velocity.  We first ask whether there is gas exceeding the escape
velocity at each radius.  With the same definition of systemic
velocity as above, we integrate the line emission that exceeds the
escape velocity to either the red or blue side of the systemic
velocity.  We then normalize by the total flux at that radius.  These
fractions are plotted as a function of radius in Figure
\ref{fig:fracesc} for the 5 objects in which at least $10\%$ of the
gas is nominally escaping for at least one radial position.  For
illustrative purposes, we focus here on the blue-shifted gas. In
addition to calculating the escaping fraction at a given radius, we
can also calculate an overall escaping fraction.  They range from
$<1\% - 25\%$ with a median value of 2\%.  

Nominally only a small fraction of the NLR gas is moving out of the
galaxy at or around the escape velocity. However, the projection
effects may be severe, and especially so because in obscured objects
the gas motions are expected to occur largely in the plane of the
sky. Therefore, our estimates are a lower limit on the actual escape
fractions (see \S 6 for details). Furthermore, as discussed further
below, we have good reason to think that the medium is clumpy.
Depending on whether the outflowing component has the same clumping
factor as the bound gas, it is difficult to translate these observed
fractions into mass fractions.

In addition to the escaping fraction, we would like to know how much
mass is involved in the outflow.  The standard method of estimating
the density of the emission line gas uses density diagnostics such as
the ratio of \sii$~\lambda 6716/\lambda 6731$ or \oii$\lambda
3729/\lambda 3726$. Neither of these is available in the Magellan
spectra, and with several hundred \kms\ velocities, the \oii\ doublet
is blended enough to be difficult to measure. The continuum-subtracted
SDSS spectra that integrate all emission within the 3\arcsec\ fiber
yield a measurement of the \sii$~\lambda 6716/\lambda 6731$ ratio for
all but the highest redshifts. Using the {\tt iraf} task {\it temden},
these can be translated into densities ranging from 250-500~cm$^{-3}$,
with a mean of 335~cm$^{-3}$. These values are consistent with those
commonly seen in spatially resolved observations of extended NLRs and
used in mass estimations (e.g.,
\citealt{nesvadbaetal2006,fustockton2009} and many others). However,
such measurements can be highly biased toward high densities in clumpy
gas. Specifically, the recombination line luminosity depends on
density as $L \propto \int {\rm d}V \alpha n_e n_p$, whereas mass goes
like $M \propto \int {\rm d}V n_p$, so the mass of the gas, its
density and degree of clumpiness and its line luminosity are related
through
\begin{equation}
M_g=1.7 \times 10^9 L_{H \beta}^{41} \langle n_e \rangle ^{-1} \kappa^{-1} M_{\sun}.
\end{equation}
Here we used a recombination coefficient 
$\alpha=1.62\times 10^{-14}$~cm$^3$~s$^{-1}$
appropriate for a 20,000\,K gas, and $\kappa=\langle n_e^2
\rangle/\langle n_e\rangle^2$ is the degree of clumpiness, which by
definition is $\ge 1$ and can be substantially greater. We have adopted 
a higher temperature than typical based on the 
\oiii$~\lambda 4363\AA$/\oiii$~\lambda 5007\AA$ line ratio.  In general, 
the ratio ranges from $\sim 0.01$ in the central regions to $\sim 0.04$ 
further out.  These ratios correspond to $T=11000-23000$K, and thus we 
adopt a temperature representative of the outer regions.

The standard method of calculating the mass involved amounts to using
this equation with $\kappa=1$ and $n_e$ of a few $\times 100$
cm$^{-3}$, and produces an absolute minimum on the gas mass visible in
the emission lines of a few $\times 10^7 M_{\sun}$. However, such high
densities are in direct conflict with our observations.  For one
thing, we see high \oiii/\hbeta\ ratios, and thus high ionization
parameters, and presumably low densities, at large radius.  Also, the
observed extended scattering regions in obscured quasars place an
independent constraint on gas densities
\citep{zakamskaetal2006}. Scattered light flux is $\propto \int {\rm
  d} V n_s$, where $n_s$ is the density of scattering particles,
electrons or dust. Assuming purely electron scattering, \hst\
observations can be fit by density profiles that decline as
$r^{-(1.5-2.5)}$ and with density $\langle n_e \rangle=1$ cm$^{-3}$ at
a distance of about 3 kpc from the center (Figure
\ref{fig:scatter}). The scattering angle is not well known, but it
introduces only about a factor of two uncertainty in this
measurement. Dust particles are even more efficient scatterers than
electrons, so in the more realistic case of dust scattering, which is
suggested by several lines of observational evidence
\citep{zakamskaetal2005}, the implied mean density is constrained to
be even smaller, $\langle n_e \rangle (1{\rm kpc})\sim 0.016$
cm$^{-3}$. The uncertainties are larger in the case of dust
scattering, because the density measurement is sensitive to the
assumed gas-to-dust ratio and the scattering angle (for this
particular value, 90\degr\ and Small Magellanic Cloud dust,
\citealt{draine2003}), but nevertheless it is clear that the scattered
light observations require much lower densities than those implied by
\sii\ ratios. The two measurements can be reconciled if the gas is
highly clumped, so that most of the luminosity is coming from
high-density clumps, whereas the mass and the scattering cross-section
are dominated by low-density gas.

While a detailed modeling of all observables is beyond the scope of
this paper, we use a toy model in which the mass of the emitting gas
at each density is a power-law function of the density, with power-law
index $-\alpha$ between $n_{\rm min}$ and $n_{\rm max}$, to estimate
the clumping factor. Since the \sii\ line ratios are usually observed
to be between the low-density and the high-density asymptotes, values
of $n_{\rm max}=$ a few times the critical density are required; we
use $n_{\rm max}=10^3$ cm$^{-3}$. At the same time, for $1<\alpha<2$
the minimal density is constrained to be $n_{\rm
  min}=(1-\alpha)\langle n_e \rangle/\alpha$ by the scattering
observations. With these constraints, the clumping factor is
\begin{equation}
\kappa=\frac{(1-\alpha)^2}{\alpha(2-\alpha)}\left(\frac{n_{\rm
        max}}{n_{\rm min}}\right)^{2-\alpha}.
\end{equation}
For example, for $\alpha=1.5$ and $\langle n_e \rangle=0.016$
cm$^{-3}$, for each $10^{41}$ erg~s$^{-1}$ of H$\beta$ emission, the
mass of the emitting gas is $M_g=7\times 10^8 
M_{\sun}$. This estimate can only be considered very approximate,
since the derived mass is quite sensitive to the specific assumptions
about clumping (for example, it varies by 2 dex as $\alpha$ varies
between 1 and 2). Nevertheless, we point out that the standard method
of mass determination likely produces an underestimate of the true mass
and that the scattering observations provide a valuable constraint on
the physical conditions in the NLR.  

In short, we see compelling evidence that the NLR is clumpy.  As a
result, it is difficult to estimate robust gas masses, and thus
difficult to determine what fraction of the gas may be expelled by
potential AGN outflows.  

\section{Two Candidate Dual Obscured Quasars}

Many recent surveys have identified potential dual active galaxies
(i.e., two active galaxies with $\sim$ kpc separations) 
as narrow-line
objects with multiple velocity peaks in the \oiii\ line in SDSS
spectra \citep[e.g.,][]{wangetal2009,liuetal2010survey,
  smithetal2010}, as well as from the DEEP2 redshift survey
\citep{gerkeetal2007,comerfordetal2009deep}.  Other candidates have
been identified based on spatially offset nuclei
\citep{barthetal2008b,comerfordetal2009cosmos}.  There are two
intriguing objects in this sample that may contain dual AGNs.

\vskip -0.5truein
\hskip 0.1in
\psfig{file=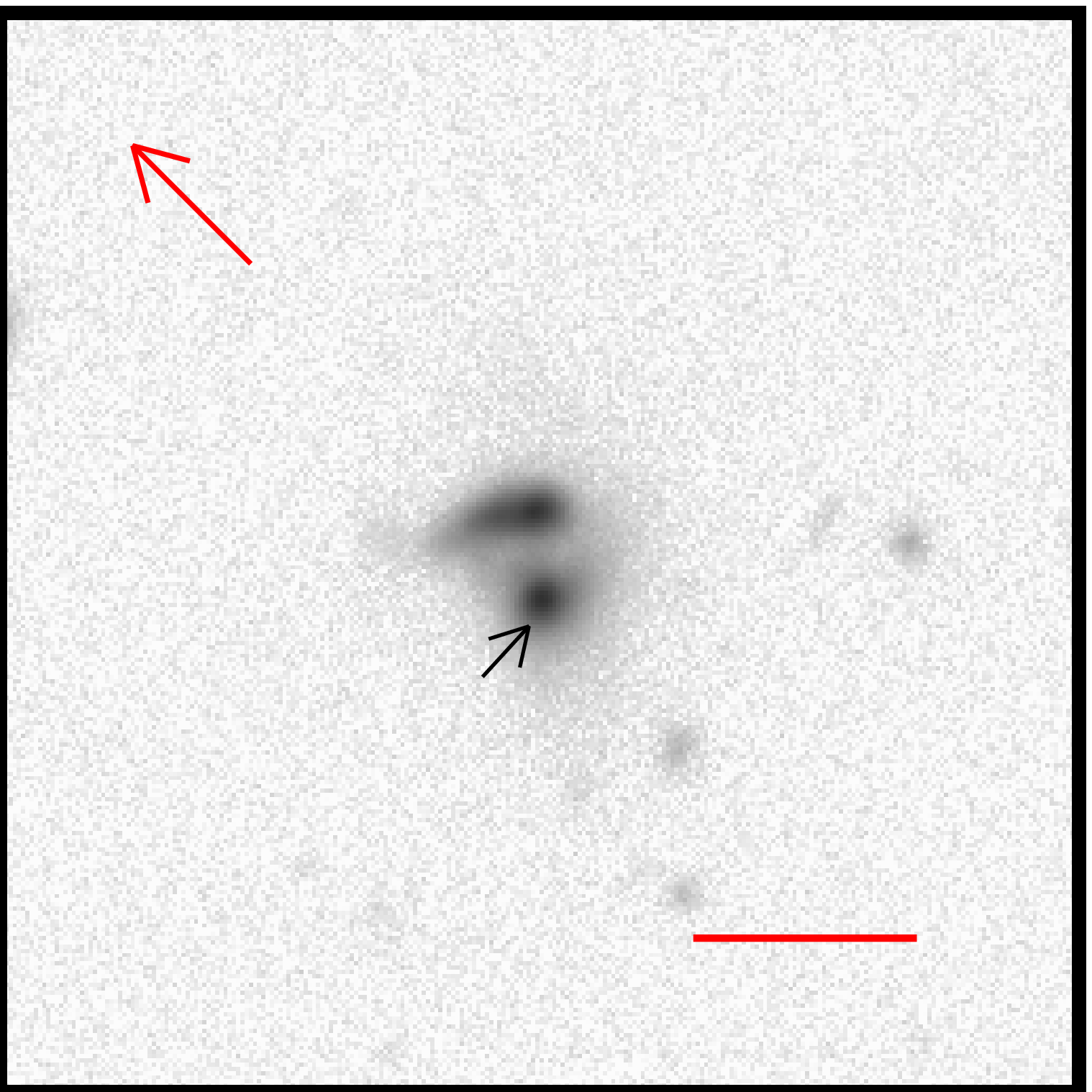,width=0.4\textwidth,keepaspectratio=true,angle=0}
\vskip -0mm \figcaption[]{ 
$r$-band image of the dual AGN SDSS~J0841+0101.  The primary object (A) is 
indicated with the black arrow, 
red arrow points North, and the scale bar is 10\arcsec\ long.  
Separation between the two AGNs is 3.8 \arcsec\ (7.6 kpc).  
\label{fig:binim}}
\vskip 5mm

The first is SDSS J1356+1026 (Fig. \ref{fig:bubbleim}), which has two
clear continuum sources, each with associated high-ionization \oiii\
emission.  Their separation is $\sim 2.5$ kpc (1\farcs1). This object
was highlighted as a potential dual AGN by both
\citet{liuetal2010survey} based on multiple velocity peaks in the SDSS
spectrum and by \citet{fuetal2010} from Keck AO imaging.  We have
recently shown that $\sim 10\%$ of the double-peaked narrow-line
candidates also have spatially resolved dual continuum sources 
\citep{liuetal2010mag}.  It seems
natural that two galaxies would contain two BHs.  On the other hand,
there well may be a single radiating BH that is illuminating all of
the surrounding gas.  Unfortunately, our long-slit spectra do not
include \oii\ or \sii, which would give us a handle on the electron
densities, and thereby whether a single ionizing source is plausible.
Given the projected separation of 2.5 kpc, if we assume that there is
a single ionizing source associated with one of the two continua, we
would expect to see the ionization parameter decrease by a factor of
$\sim 6$ between the two targets.  In fact, the \oiii/\hbeta\ ratios
are within $10\%$ of each other, as are the \oiii\ fluxes.  On the other
hand, the very high ionization parameter seems to extend over the
entire nebulosity ($\sim 10$ kpc).  Of course, the accreting BH may
sit between the two continuum sources.  Definitive proof requires the
detection of X-ray or radio cores associated with each continuum
source.

SDSS J0841+0101 shows much less ambiguous evidence for a pair of
accreting BHs, with a projected separation of 3\farcs8 (7.6 kpc;
Fig. \ref{fig:binim}).  It would not be included in double-peaked
samples assembled from the SDSS because the separation on the sky
between the two components is larger than the SDSS 3\arcsec\ fibers.
Nevertheless, the component separations are comparable to those in the
Liu et al. sample.  \citet{liuetal2010mag} show that the double-peaked
samples are probably dominated by single AGNs.  These observations
highlight that we are likewise missing dual AGNs with slightly larger
separations.

\begin{figure*}
\vbox{ 
\vskip -0.3truein
\hskip 0.5in
\psfig{file=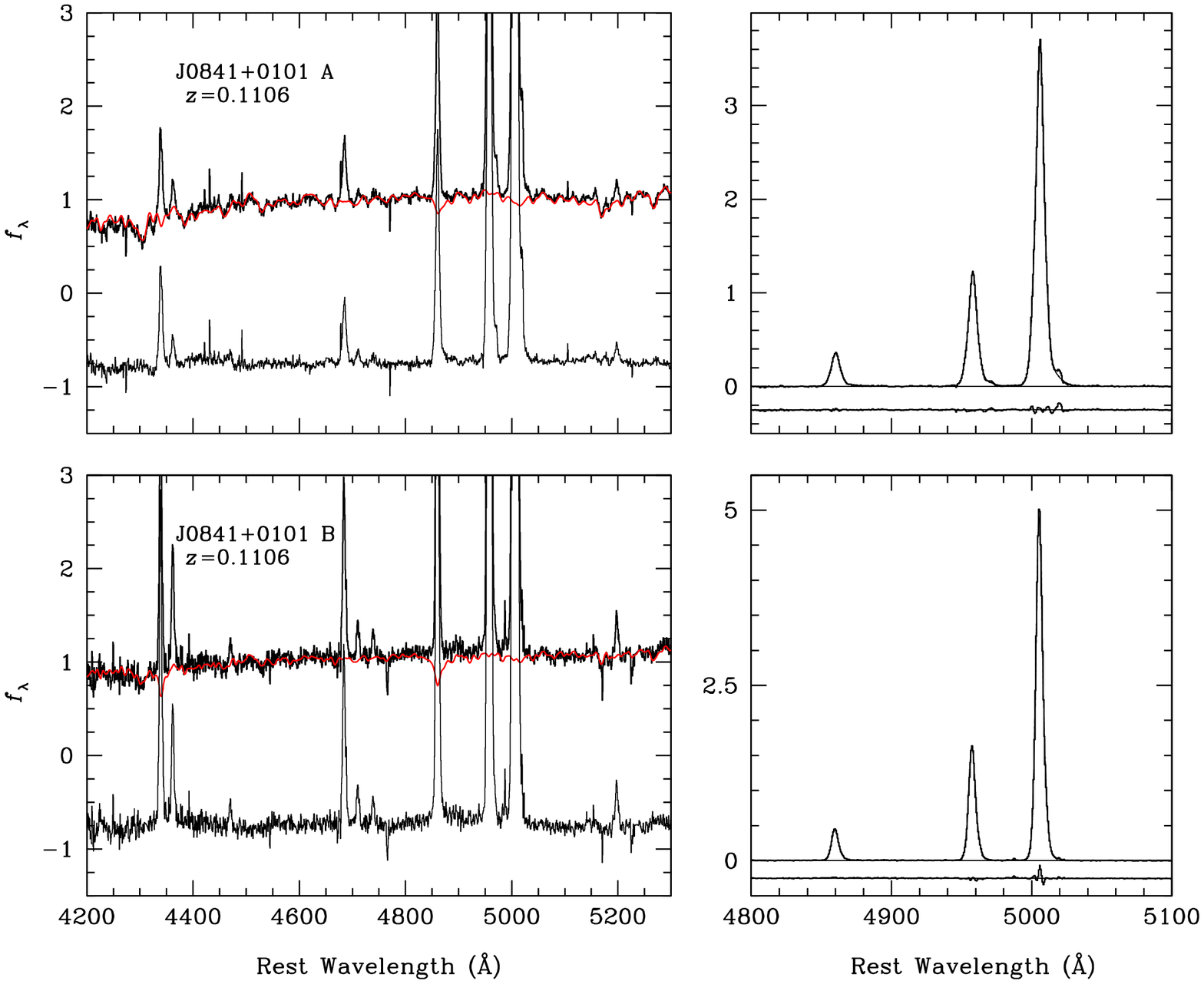,width=0.65\textwidth,keepaspectratio=true,angle=90}
}
\vskip -0mm
\figcaption[]{
The binary narrow-line quasar serendipitously discovered in our spectroscopic
campaign.  Component A was observed by the SDSS and is the more massive 
of the two systems.  
{\it Left}: Stellar velocity dispersion fits 
using the best weighted fit of an F2, K1, and K4 star. These fits 
provide both a measurements of \sigmastar\ and also a continuum subtraction.
Flux densities are normalized to the continuum. 
{\it Right}: Fits to the \hbeta\ and [O {\tiny III}] emission lines using 
multi-component Gaussians; see text for details. 
\label{fig:binspec}}
\end{figure*}

As is apparent from Figure \ref{fig:binspec}, the two AGNs are
strikingly similar in spectroscopic properties.  The \oiii\
luminosities ($\approx 10^{42}$ erg~s$^{-1}$) agree within $<0.1$ dex,
and the \oiii/\hbeta\ ratios ($\sim 10$) agree within $5\%$.  The only
clear difference is in the linewidths.  The primary galaxy (A) has a
\fwoiii$=430$~\kms, while the companion AGN (B) is narrower, with
\fwoiii$=330$~\kms.  This difference most likely reflects the fact
that A, with a stellar velocity dispersion of $\sigmastar=214 \pm
29$~\kms\ is more massive than B, with $\sigmastar=150 \pm 40$~\kms.
Taken at face value, this difference in dispersions corresponds to a
difference of a factor of nearly 10 in BH mass between the two
galaxies.  Accordingly, if the \oiii\ luminosity tracks the bolometric
luminosity, then apparently B is accreting 10 times closer to its
Eddington limit than A.  

Alternatively, there may be only a single radiating black hole. 
If there is only one quasar -- in galaxy A -- then we consider two
scenarios. The first is that the quasar in A is unobscured as seen
from B, so that the galaxy B is photoionized by the central engine in
A.  If we assume that most of the NLR emission in A is produced at
a distance $\la 1$ kpc from the nucleus, then in order to preserve the
ionization parameter (as evidenced by the similar spectra of A and B),
the difference in electron density between the two galaxies would have
to be a factor of $\ga 60$. While the dust particles in galaxy
B may scatter quasar spectrum, this emission cannot dominate the
observed spectrum (otherwise we would see a broad-line AGN in source
B). The resulting estimates of the emerging equivalent width of the
emission lines suggest that this scenario is possible, but has to be
quite tuned in order to fit observations. The second scenario is that
galaxy B is located along the obscured direction, just like the
observer, but scatters some of the A's \oiii\ emission. However, in
this scenario the ratio of [OIII] fluxes of B and A corresponds to the
fraction of photons that B intercepts, $\sim (2/7.6)2 / 4 \pi =
0.006$, contrary to the observed similarity of fluxes. In conclusion,
the picture of a single active black hole producing two objects with
similar fluxes and ionization parameters appears unlikely.

\section{Discussion \& Summary}

We are looking for direct signs of feedback in the two-dimensional
spatial extents and kinematics of the NLRs of a sample of luminous
obscured active galaxies.  Our conclusions are mixed.
On the one hand, we see clear evidence that the AGN is stirring up the
galaxy ISM.  On the other hand, we do not see signs of
galaxy-scale winds at high velocities.  However, as we argue below,
perhaps this is unsurprising.

We see two distinct signatures of a luminous accreting BH on the 
ionized gas in these galaxies.  The NLRs are much more extended at 
these high luminosities than in lower-luminosity Seyfert galaxies.  
In fact, the AGNs are effectively photoionizing gas throughout the entire 
galaxy.  This alone means that the AGN is heating
the ISM on galaxy-wide scales.  The impact of the AGN is more directly 
seen in the kinematics.  We see very few ordered radial velocity curves; instead 
the velocity distributions are typically quite flat even at large radius. 
Perhaps even more striking is that the gas velocity dispersions are
high out to large radius.  As we have argued, not only do inactive galaxies 
uniformly show a drop in gas (and stellar) velocity dispersion at large 
radius, but even in ultra-luminous infrared galaxies the gas velocity 
dispersions are observed to drop at large radius.  We cannot therefore 
attribute the gas stirring to gravitational effects such as mergers.  It 
is most natural to implicate the accreting BH.  

On the other hand, overall the velocities we observe in the NLR gas
are not very high (a few hundred km~s$^{-1}$).  Taken at face value,
our crude estimates suggest that very little of the ISM is moving fast
enough to escape the galaxy, although a clumpy NLR complicates our
ability to estimate this fraction robustly.  In only one case do we
see the spectacular outflowing nebulosity one might imagine in
thinking of AGN feedback (SDSS~J1356+1026).  Before we can rule out that
any gas is unbound from these galaxies, however, we should consider
the impact of projection effects, potential observational biases, and
some theoretical expectations.

Our observations suggest that ionized gas is ubiquitous within the
galaxy but rare at larger (e.g., 10 kpc) scales.  As explained above,
the observed ratio of obscured to unobscured objects leads us to
assume an ionization cone opening angle of $\sim 120\degr$.  With such
a large opening angle, we would expect our slit to intercept the NLR
nearly all the time, as we observe.  On the other hand, we see
extended gas on 10 kpc scales in only one case.  Furthermore, the
\hst\ continuum images show extended emission on these large scales,
but with a much smaller opening angle of 20-60\degr.  Similarly, we
have visually inspected the most luminous obscured AGNs from the Reyes
et al. sample with $0.16<z<0.3$ and found evidence for small opening
angles from the broad-band images (which have significant \oiii\ light
in the $r$ band).  Probably we are seeing the effects of surface
brightness dimming at the outer reaches of the bicone.  Although the
true opening angle is large (120\degr), only a much narrower inner
cone can be observed at 10 kpc.  Taking the smaller opening angles, we
expect to see extra-galactic extended gas only 20-40\% of the time.
That fraction is not inconsistent with the number of objects that we
observe with emission line regions extending beyond their host
galaxies.

\begin{figure*}
\vbox{ 
\vskip +2.5truein
\hskip 0.3in
\psfig{file=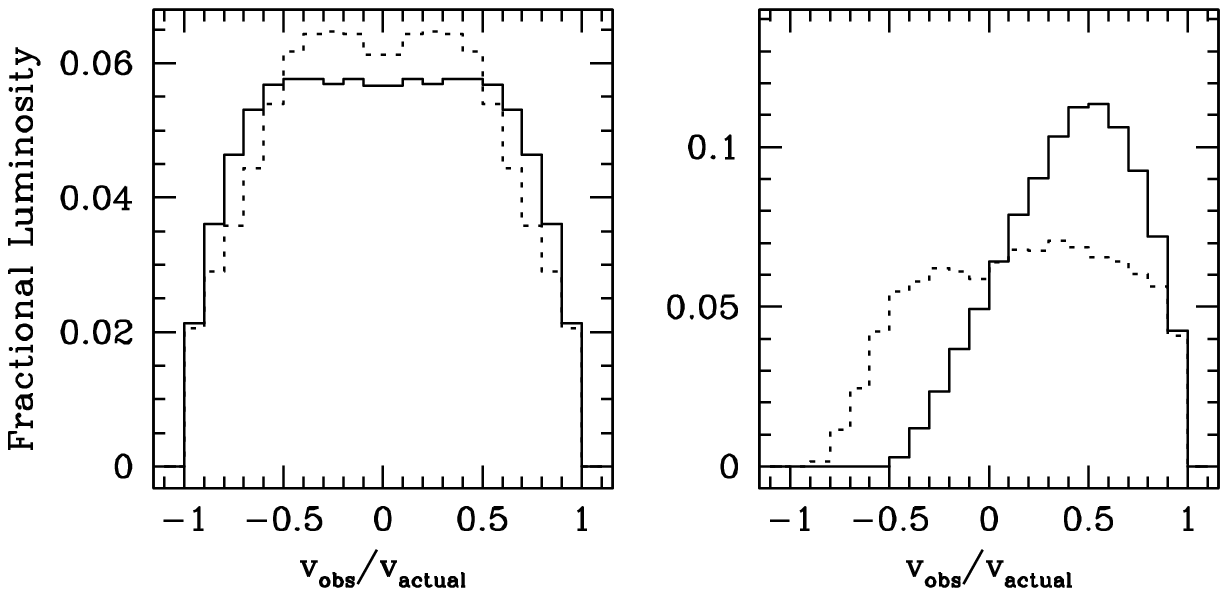,width=0.8\textwidth,keepaspectratio=true,angle=0}
}
\vskip -0mm \figcaption[]{ 
  To illustrate the magnitude of the projection effects, we present
  the results of two Monte Carlo simulations of observed velocity
  distributions for a conical outflow with an opening angle of
  60\degr\ ({\it solid histogram}) and 120\degr\ ({\it dotted
    histogram}). The line of sight is constrained to fall outside of
  the cone but is otherwise drawn from the appropriate random
  probability function. The outflow is assumed to emit uniformly and
  to have a constant outflow velocity $v_{\rm actual}$, but we measure
  the (smaller) radial velocity $v_{\rm obs}$. On the left, we show
  the distribution of velocities from both bicones, while on the right
  we show only the approaching bicone. Because many of the streamlines
  of the gas lie close to the plane of the sky, the observed
  velocities are biased to be significantly smaller than the real
  ones. This figure demonstrates that even if we detect only a small
  fraction of the gas at high velocities approaching the escape speed,
  a large fraction of the gas may actually be escaping.
\label{fig:velproject}}
\end{figure*}

Projection effects also preferentially bias us against detecting the
true outflowing velocities.  These are obscured objects, and on large
scales we see evidence for ionization cones in the \hst\ continuum
imaging.  We thus expect the largest accelerations to occur in the
plane of the sky.  We perform a Monte Carlo simulation in which the
NLR is modeled as a biconical outflow with constant velocity as a
function of radius, assuming different opening angles for the bicone
(Fig. \ref{fig:velproject}).  We sample random lines of sight outside
of the bicone, and find that while the intrinsic velocity is uniformly
high, we only expect to observe high (e.g., approaching escape)
velocities a small fraction of the time.  

These simulations take into account only the bias introduced by
projection effects and assume constant velocity and uniform emissivity
within the bi-cone. We also considered more realistic models, in which
velocity varies as a function of distance ($-2<d \log v/ d \log r<1$) 
from the center, mass conservation is satisfied and the
emissivity correspondingly declines as $n^2(r)$. Due to the decline of
emissivity in these models, at a projected distance $d$ from the
center the observed brightness is dominated by the location physically
closest to the center (that is, $r=d$), and this gas is moving exactly
in the plane of the sky exacerbating the projection bias. In the case
$v(r) \propto r^{-2}$ the emissivity may be uniform, but the integral
along the line of sight is dominated by gas that moves slower than the
gas at $d$ because of the declining velocity profile. Therefore, in
these more realistic situations we find a radial velocity distribution
that is more peaked at zero than shown in
Fig. \ref{fig:velproject}. Thus, while we observe small escaping
fractions, once projection effects are accounted for, the observations
may be consistent with high velocities in a large fraction of the
gas.

Finally, there is the possibility that the outflows operate
predominantly on small scales. In local low-luminosity sources
outflows are observed only within the inner hundreds of pc
\citep[e.g.,][]{crenshawkraemer2000}.  In addition, recent simulations
by \citet{debuhretal2010} suggest that BHs do self-regulate their own
growth but do not generate galaxy-wide outflows.

Of course, other factors may be at play as well.  There is the
possibility that some fraction of the ionizing photons have escaped
the galaxy \citep[e.g.,][]{netzeretal2004}, or even that we are seeing
galaxies in some pre-outburst phase, as may be expected if obscured
accretion tends to accompany the late stages of merging and star
formation activity \citep[e.g.,][]{mihoshernquist1994,
sandersmirabel1996,canalizostockton2001}.

It is interesting to compare with simulations of galaxy-scale
outflows.  We start with the work of \citet{progaetal2008} \citep[see
also][]{proga2007,kurosawaproga2009}.  These simulations focus on
smaller scales than those we probe, extending no further than 10 pc.
However, it is at least a starting point for comparison.  The
simulations include radiative heating by both an accretion disk and an
X-ray corona, and look at the impact of varying the density and
temperature structure, as well as rotation, of the gas.  We highlight
a few generic conclusions from their studies that are very relevant to
our work.  First of all, the final flow includes both an equatorial
inflow and a bipolar outflow.  Consistent with our work, the opening
angle of the outflowing cone can be quite wide (up to 160\degr).  Also
interesting to note is that the outflows can be dynamic, clumpy as we
observe, and with multiple temperatures (ranging from the 10$^4$ K gas
observed here all the way to X-ray emitting temperatures).  It remains
to be seen whether the outflows on pc scales will propogate to larger
(galaxy-wide scales).

A recent study by Hopkins et al. (in preparation) of outflows driven
by AGNs in numerical simulations demonstrates several surprising
similarities to the kinematics of the ionized gas we see in our
study. The observations suggest that outflows are clumpy because the
measurements of rms density and the mean density are highly
discrepant. The simulations suggest that outflows are clumpy because
they are subject to Rayleigh-Taylor instabilities. Furthermore, the
rate of the decline of mean density with distance from the center seen
in scattering observations is similar to that seen in numerical
simulations where the motion of the gas becomes ballistic at large
distances. The masses and the velocities of the outflows that we find
are quite similar to those seen in numerical simulations, and although
the kinetic energies of the outflows ($\sim 10^{42}$ erg~s$^{-1}$) are
just a small fraction of the total energy output of the AGN, the
simulations suggest that the wind is in fact driven by a much stronger
coupling of the AGN output to the gas. The small kinetic energies that
we see at this late ($\sim 10^7$ years) stage are simply left-overs
after much of the energy was efficiently radiated by the
outflow. While these qualitative similarities are very encouraging,
the specific mechanism responsible for coupling of the black hole
output to the gas on much smaller spatial scales (which then develops
into the relic outflow we see now) remains unidentified.

In short, it is clear that the presence of the AGN at the galaxy
center impacts the entire galaxy.  Whether significant mass outflows
are driven, particularly in the radio-quiet regime considered here,
remains an open question.  The next step for this type of analysis is
already underway.  Integral-field unit observations
\citep[e.g.,][]{villarmartinetal2010}, particularly with a wider
wavelength coverage, will remove some of the ambiguities we struggle
with.

\acknowledgements
We thank G. Novak for numerous interesting discussions, and P. Hopkins 
for sending us a manuscript in advance of publication. We thank the 
referee, Sylvain Veilleux, for a very prompt, careful and helpful report that 
significantly improved this manuscript.
Research by A.J.B. is supported by NSF grant AST-0548198.

\appendix
This appendix includes all of the two-dimensional information for 
all galaxies that are spatially resolved in our observations (Figs. 12-20).  Note in 
particular the high velocities and dispersions at large radius.

\begin{figure*}
\vbox{ 
\vskip -0.75truein
\hskip -0.2in
\psfig{file=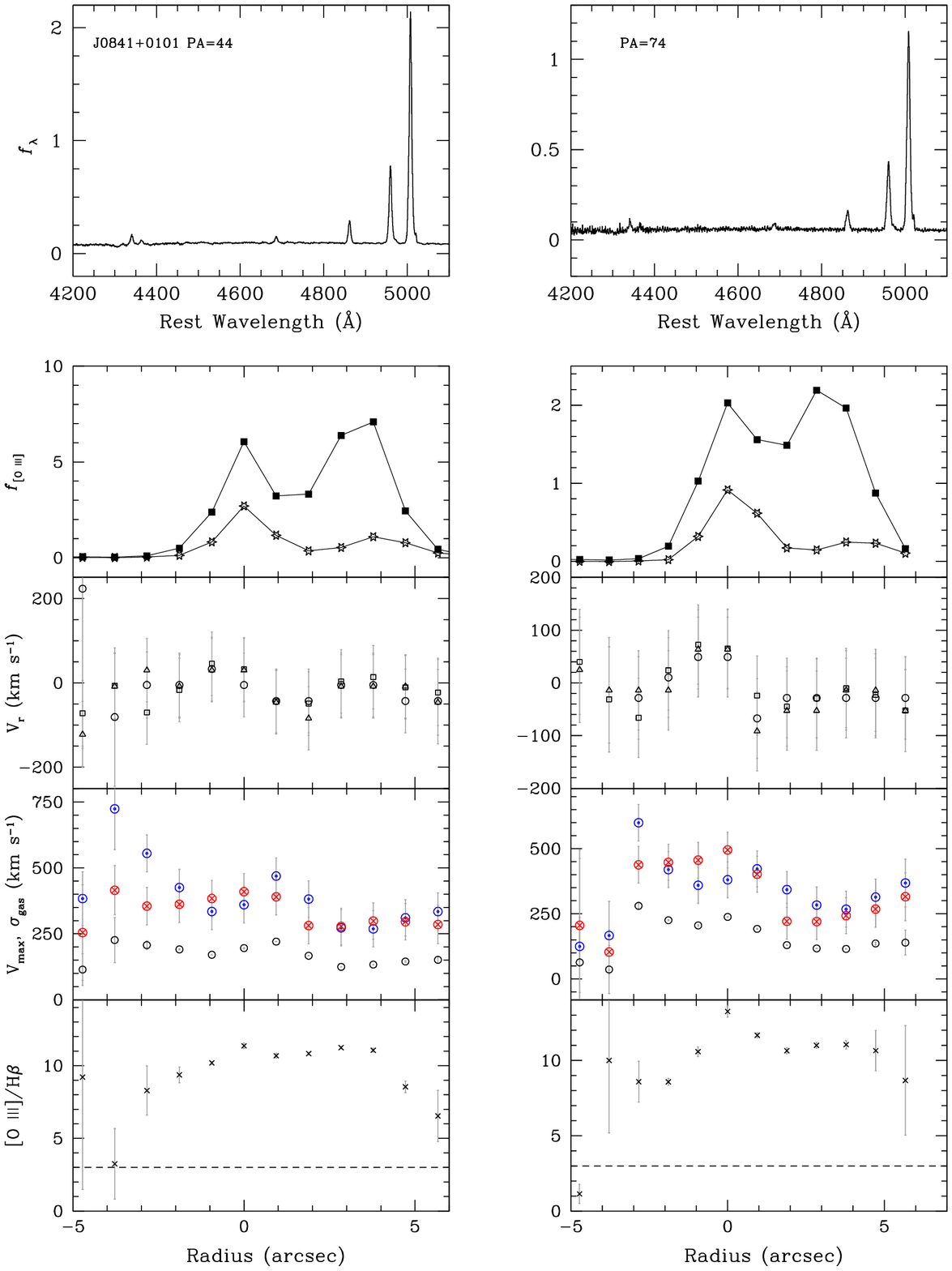,width=0.85\textwidth,keepaspectratio=true,angle=0}
}
\vskip -0mm
\figcaption[]{
Resolved spectroscopic measurements for SDSS~J0841+0101.  
In the top panel we show the 
central spectrum, extracted within the inner $\sim 1\arcsec$ arbitrarily 
normalized.  The next
panel shows the flux in the [O {\tiny III}] line ({\it filled black squares}) and 
in the continuum ({\it open stars}) as a function of radius 
(10$^{-15}$~erg~cm$^{-2}$~s$^{-1}$). 
The velocity panel shows the velocity offset between [O {\tiny III}] and  the 
``systemic'' velocity as measured from \hbeta\ in the nuclear spectrum 
({\it filled black circles}), 
the velocity dispersion in the gas ({\it open black circles}), 
and the maximum velocity (measured at $20\%$ of the line maximum) 
to the red ({\it red open circle with cross}) and to the blue 
({\it blue open circle with dot}).  Finally, the ratio of 
[O {\tiny III}]/\hbeta\ is plotted ({\it black crosses}).  A ratio of [O {\tiny III}]/\hbeta\ of three 
is noted with the dashed line.
\label{twod0841}}
\end{figure*}

\begin{figure*}
\vbox{ 
\vskip -0.75truein
\hskip -0.2in
\psfig{file=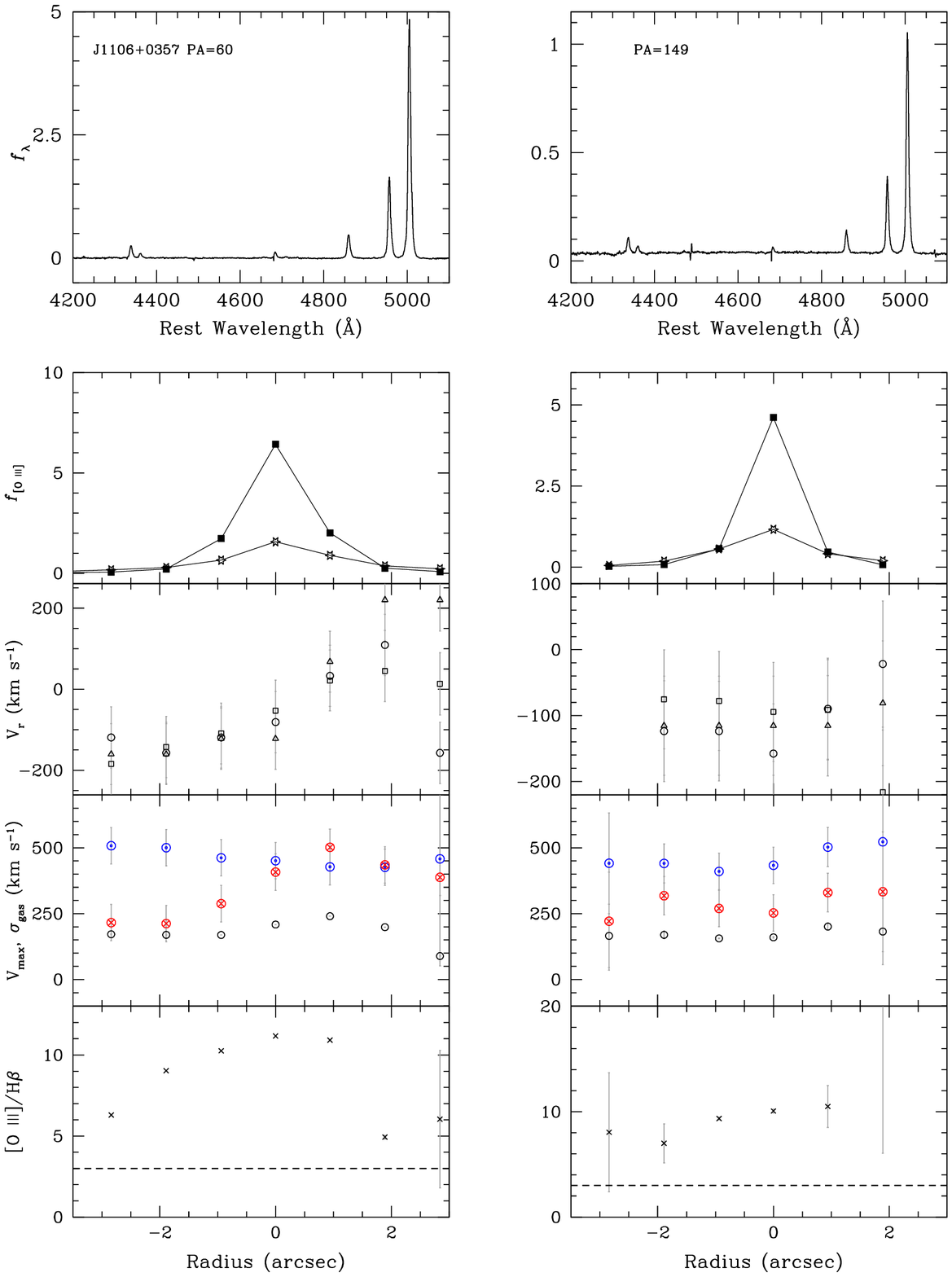,width=0.85\textwidth,keepaspectratio=true,angle=0}
}
\vskip -0mm
\figcaption[]{
Resolved spectroscopic measurements for SDSS~J1106+0357.
Symbols as above.
\label{twod1106}}
\end{figure*}

\begin{figure*}
\vbox{ 
\vskip -0.75truein
\hskip -0.2in
\psfig{file=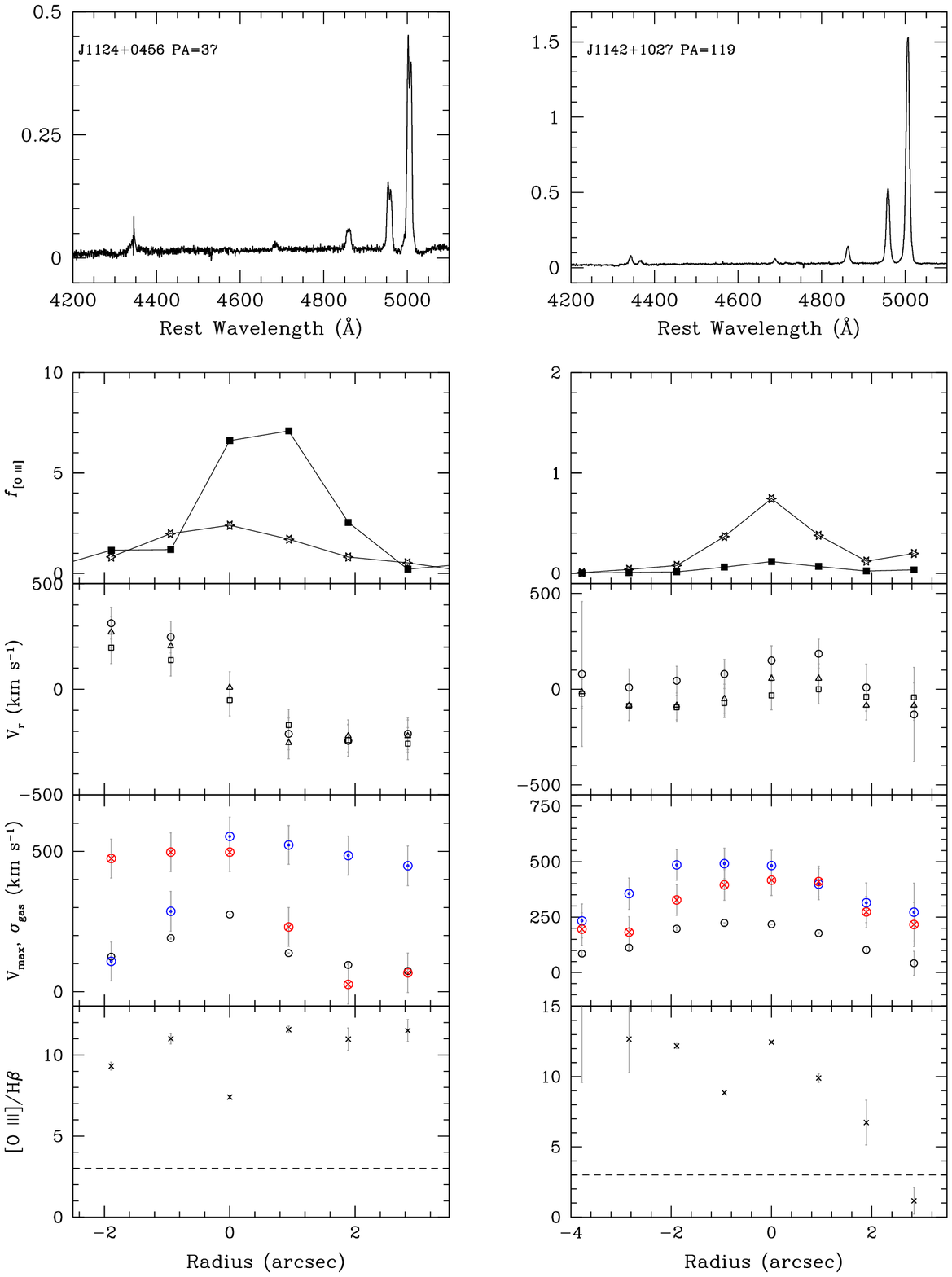,width=0.9\textwidth,keepaspectratio=true,angle=0}
}
\vskip -0mm
\figcaption[]{
Resolved spectroscopic measurements for SDSS~J1124+0456 and SDSS~J1142+1027.
Symbols as above.
\label{twod1106}}
\end{figure*}

\begin{figure*}
\vbox{ 
\vskip -0.75truein
\hskip -0.2in
\psfig{file=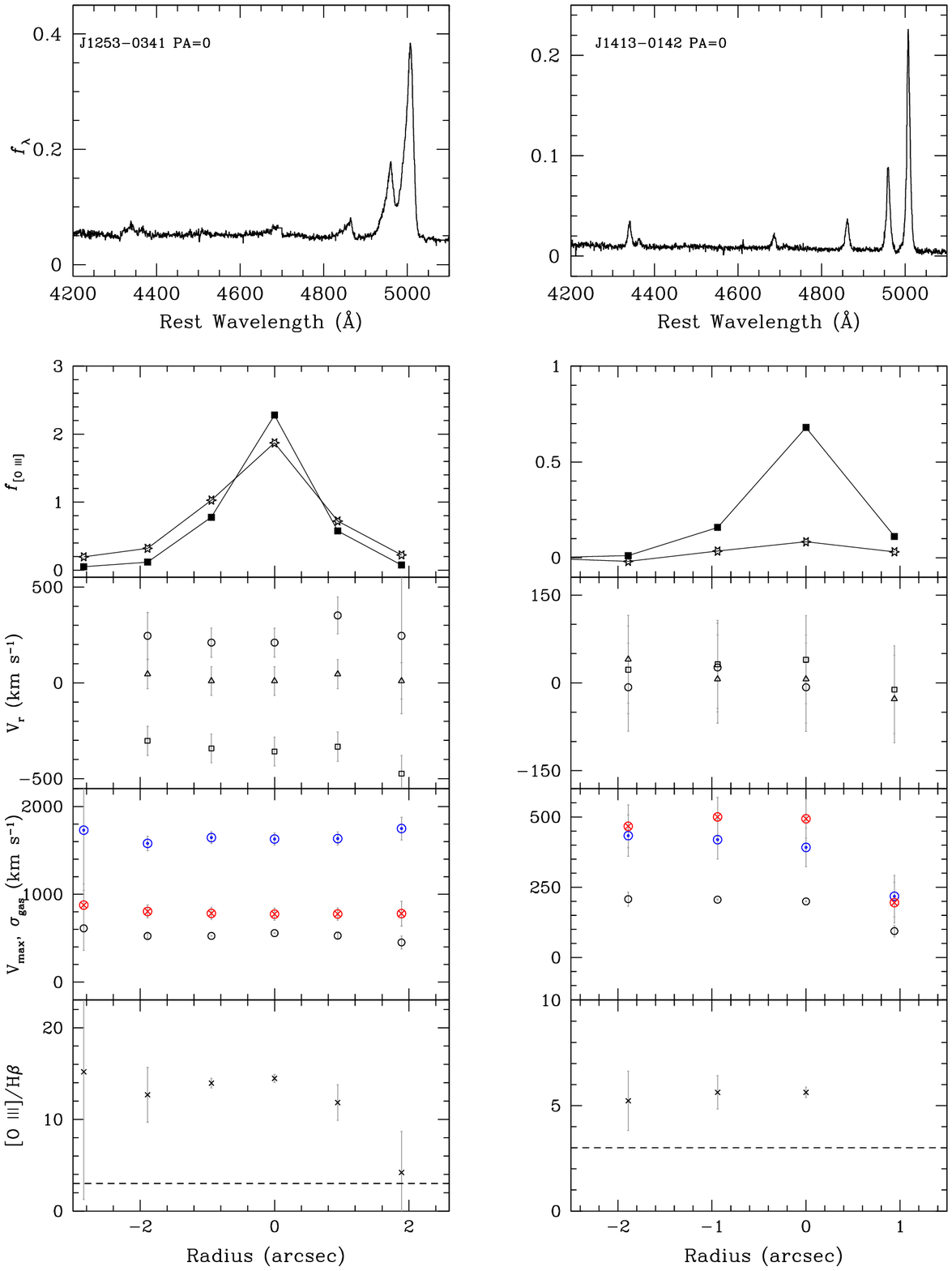,width=0.9\textwidth,keepaspectratio=true,angle=0}
}
\vskip -0mm
\figcaption[]{
Resolved spectroscopic measurements for SDSS~J1253$-0341$ and 
SDSS~J1413$-0142$. Symbols as above.
\label{twod1106}}
\end{figure*}

\begin{figure*}
\vbox{ 
\vskip -0.75truein
\hskip -0.2in
\psfig{file=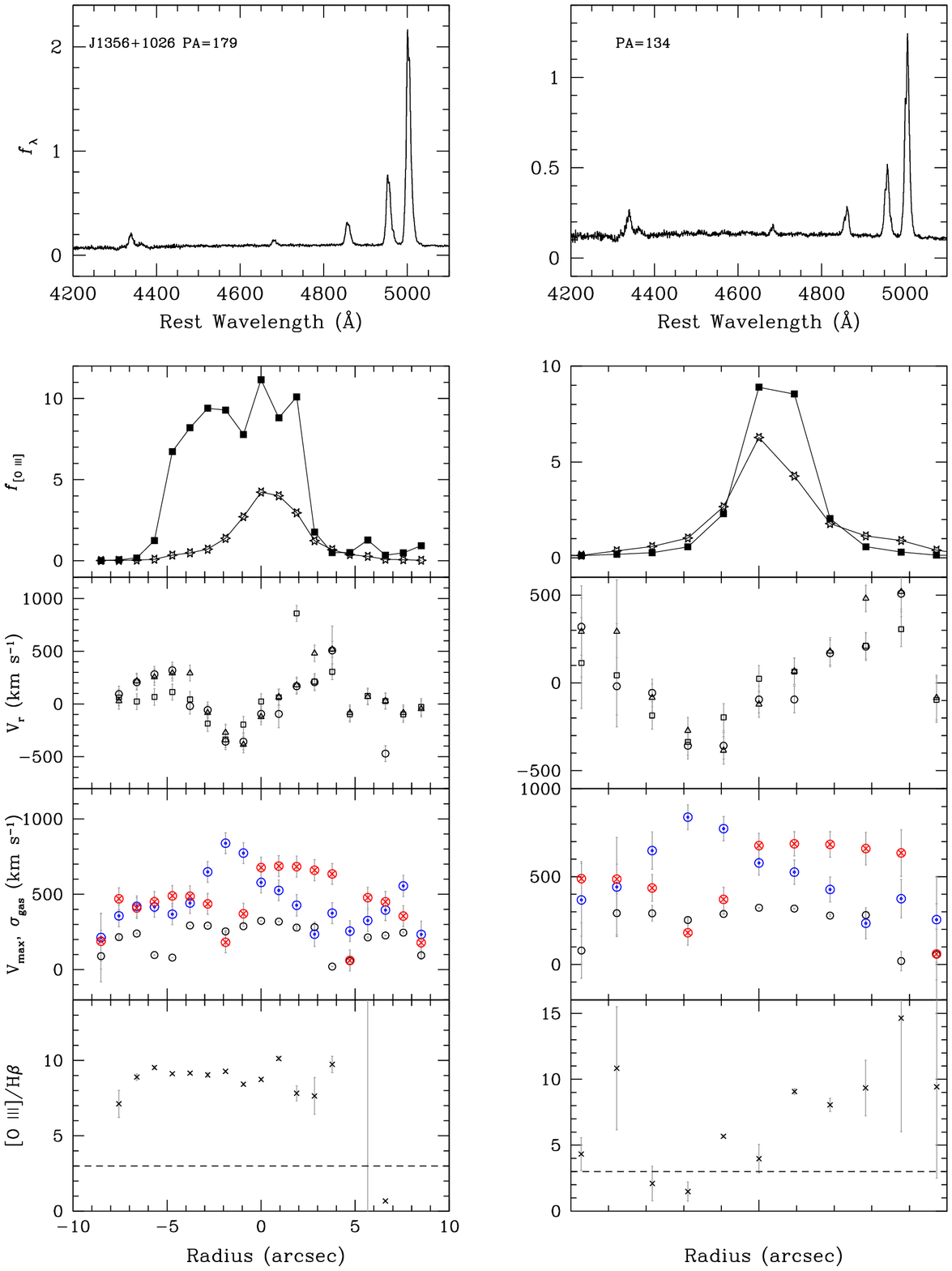,width=0.9\textwidth,keepaspectratio=true,angle=0}
}
\vskip -0mm
\figcaption[]{
Resolved spectroscopic measurements for SDSS~J1356+1026.  Symbols as above, shown 
for each slit position.
\label{twod1356}}
\end{figure*}

\begin{figure*}
\vbox{ 
\vskip -0.75truein
\hskip -0.2in
\psfig{file=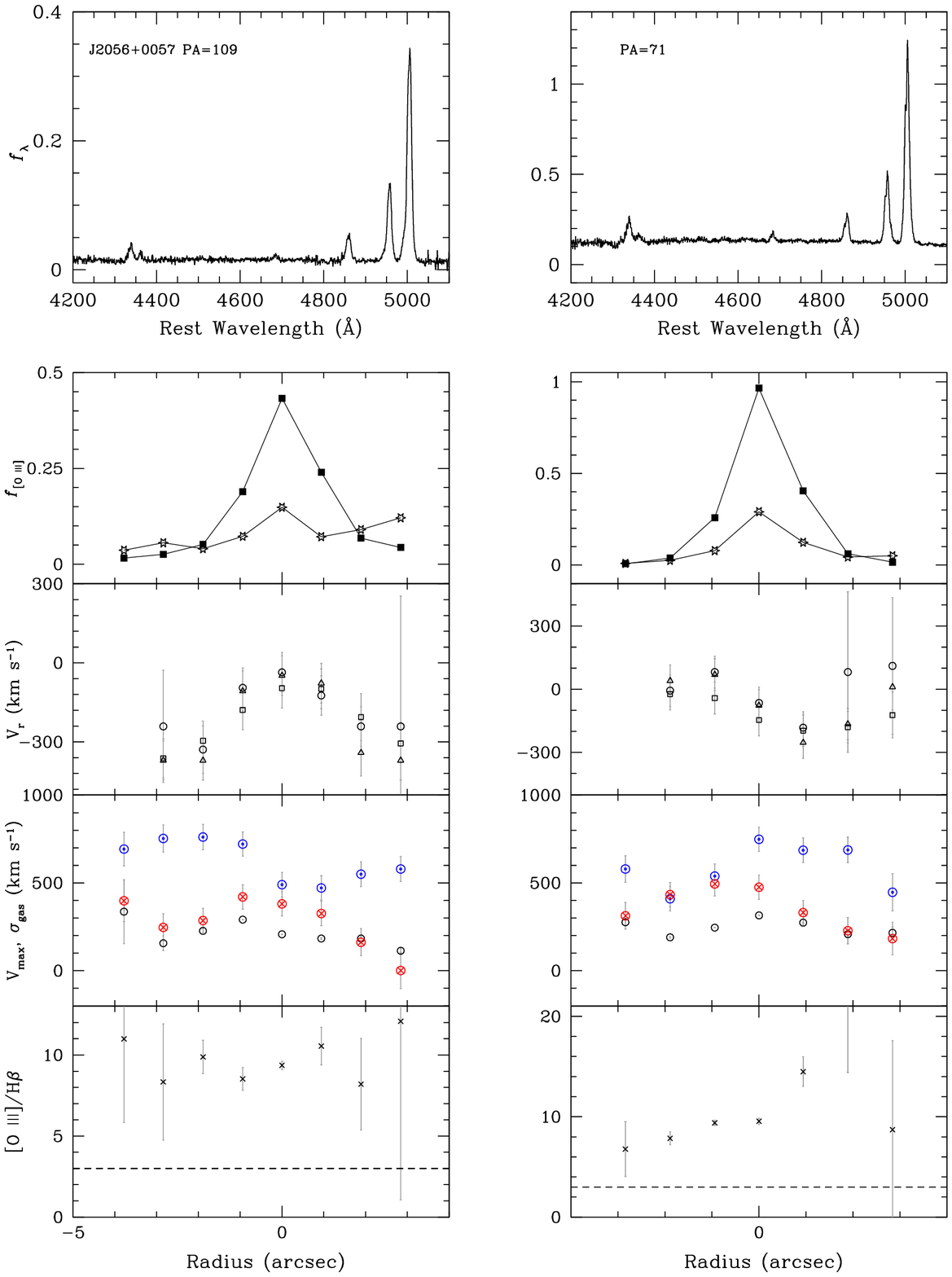,width=0.9\textwidth,keepaspectratio=true,angle=0}
}
\vskip -0mm
\figcaption[]{
Resolved spectroscopic measurements for SDSS~J2056+0057.  Symbols as above, shown 
for each slit position.
\label{twod2056}}
\end{figure*}

\begin{figure*}
\vbox{ 
\vskip -0.75truein
\hskip -0.2in
\psfig{file=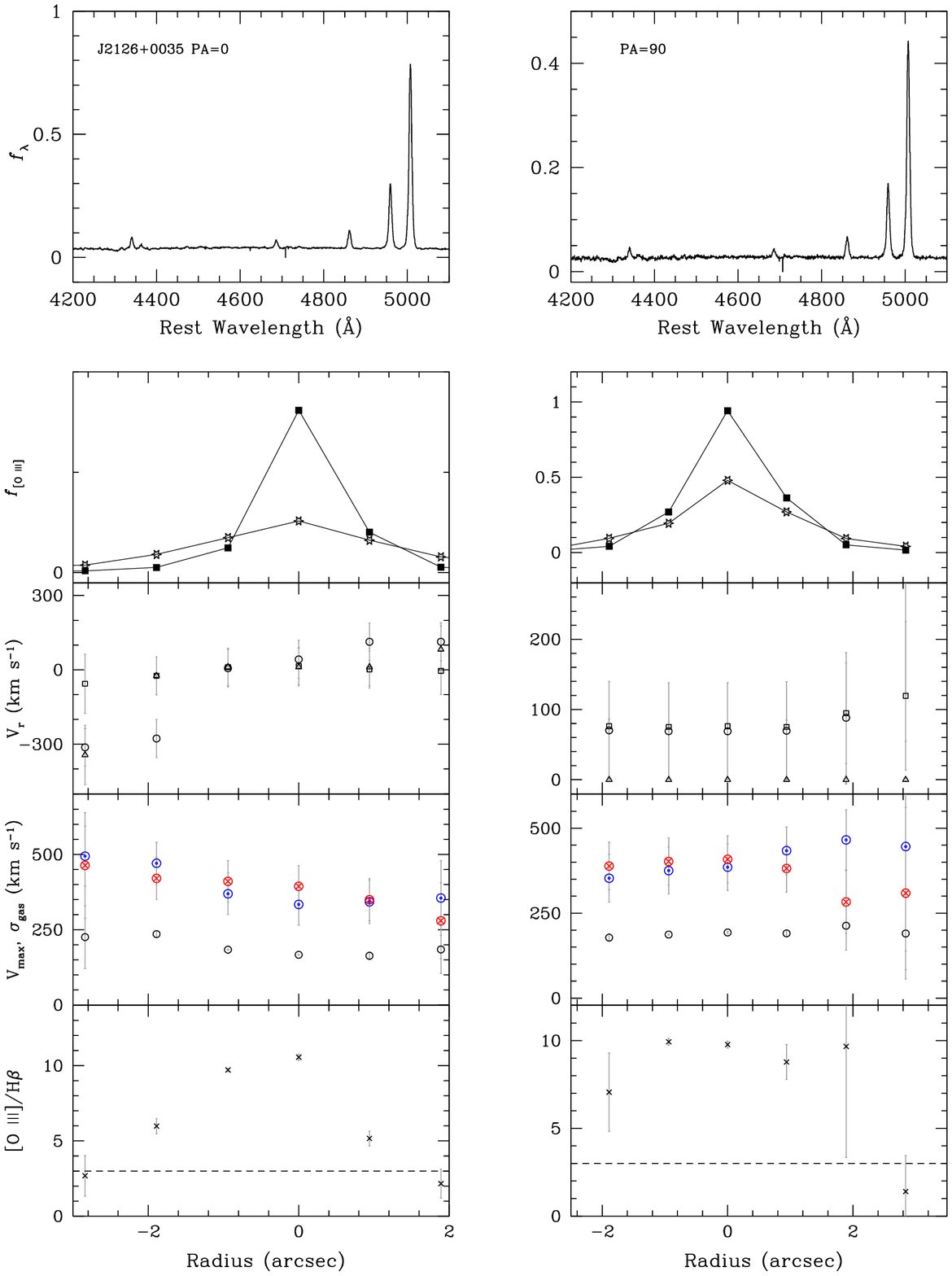,width=0.9\textwidth,keepaspectratio=true,angle=0}
}
\vskip -0mm
\figcaption[]{
Resolved spectroscopic measurements for SDSS~J2126+0035.  Symbols as above, shown 
for each slit position.
\label{twod2126}}
\end{figure*}

\begin{figure*}
\vbox{ 
\vskip -0.75truein
\hskip -0.2in
\psfig{file=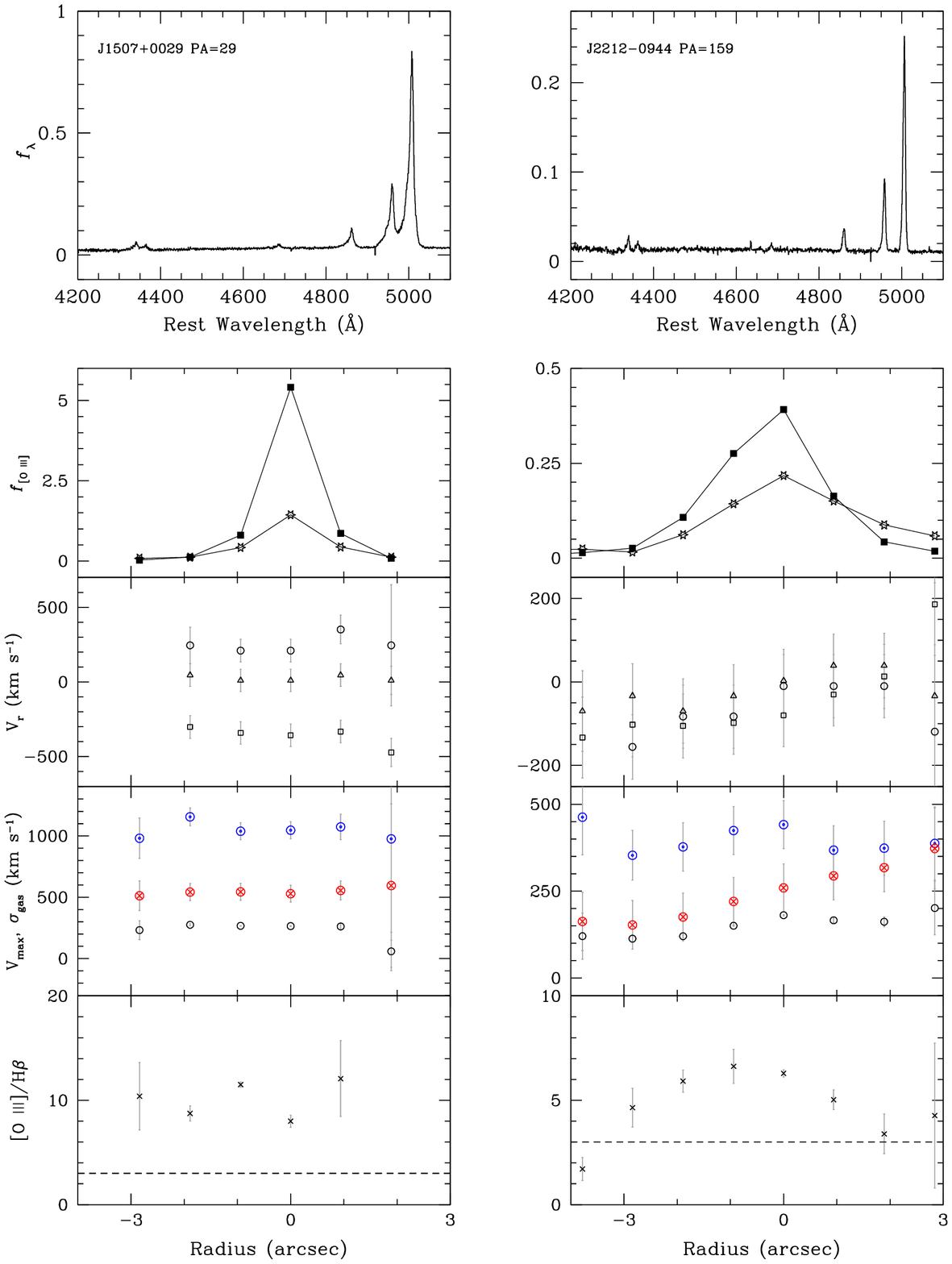,width=0.9\textwidth,keepaspectratio=true,angle=0}
}
\vskip -0mm
\figcaption[]{
Resolved spectroscopic measurements for SDSS~J1507+0029 and 
SDSS~J2212$-0944$.  Symbols as above.
\label{twod2212}}
\end{figure*}

\begin{figure*}
\vbox{ 
\vskip -0.75truein
\hskip -0.2in
\psfig{file=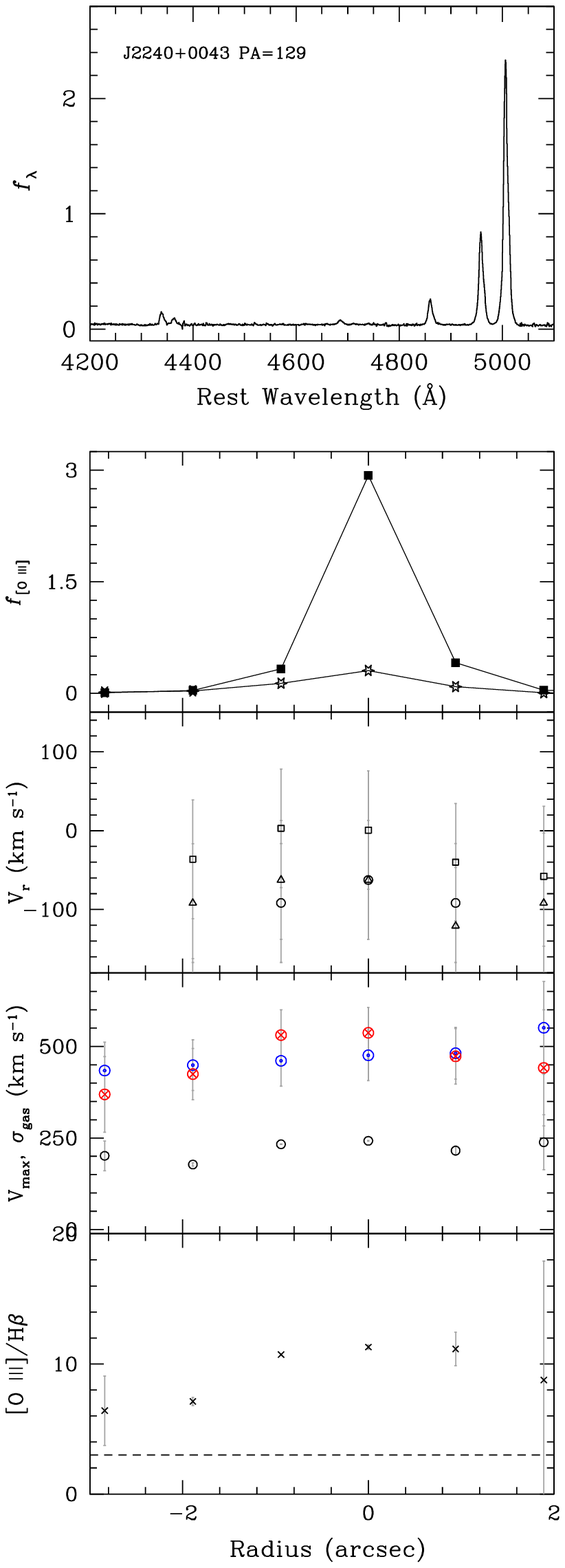,width=0.4\textwidth,keepaspectratio=true,angle=0}
}
\vskip -0mm
\figcaption[]{
Resolved spectroscopic measurements for SDSS~J2240+0043.  
Symbols as above.
\label{twod2212}}
\end{figure*}

\end{document}